%% file: main.tex
\documentclass[reprint, amsmath,amssymb, aps, pra]{revtex4-2}

\usepackage{graphicx}
\usepackage{dcolumn}
\usepackage{bm}
\usepackage{xcolor}
\usepackage{textcomp}

\begin{document}

\preprint{APS/123-QED}

\title{Quantum error correction below the surface code threshold}

\author{Google Quantum AI and Collaborators}\noaffiliation

\date{\today}

\begin{abstract}
Quantum error correction~\cite{shor1995scheme, gottesman1997stabilizer, dennis2002topological, kitaev2003fault} provides a path to reach practical quantum computing by combining multiple physical qubits into a logical qubit, where the logical error rate is suppressed exponentially as more qubits are added.
However, this exponential suppression only occurs if the physical error rate is below a critical threshold.
In this work, we present two surface code memories operating below this threshold: a distance-7 code and a distance-5 code integrated with a real-time decoder.
The logical error rate of our larger quantum memory is suppressed by a factor of $\Lambda = 2.14 \pm 0.02$ when increasing the code distance by two, culminating in a 101-qubit distance-7 code with 0.143\% $\pm$ 0.003\% error per cycle of error correction.
This logical memory is also beyond break-even, exceeding its best physical qubit's lifetime by a factor of $2.4 \pm 0.3$.
We maintain below-threshold performance when decoding in real time, achieving an average decoder latency of 63\,\textmu s at distance-5 up to a million cycles, with a cycle time of 1.1\,\textmu s. 
To probe the limits of our error-correction performance, we run repetition codes up to distance-29 and find that logical performance is limited by rare correlated error events occurring approximately once every hour, or $3$$\times$$10^9$ cycles.
Our results present device performance that, if scaled, could realize the operational requirements of large scale fault-tolerant quantum algorithms.

\end{abstract}

\maketitle

\section{Introduction}

Quantum computing promises computational speedups in quantum chemistry \cite{aspuru2005simulated}, quantum simulation \cite{lloyd1996universal}, cryptography \cite{shor1999polynomial}, and optimization \cite{farhi2001quantum}.
However, quantum information is fragile and quantum operations are error-prone. 
State-of-the-art many-qubit platforms have only recently demonstrated $99.9\%$ fidelity entangling gates \cite{mckay2023benchmarking, decross2024computational}, far short of the $<10^{-10}$ error rates needed for many applications \cite{campbell2021early, kivlichan2020improved}.
Quantum error correction is postulated to realize high-fidelity logical qubits by distributing quantum information over many entangled physical qubits to protect against errors.
If the physical operations are below a critical noise threshold, the logical error should be suppressed exponentially as we increase the number of physical qubits per logical qubit.
This behavior is expressed in the approximate relation
\[
\varepsilon_d \propto \left(\frac{p}{p_\text{thr}}\right)^{(d + 1) / 2} \tag{1}
\]
for error-corrected surface code logical qubits~\cite{kitaev2003fault, dennis2002topological, fowler2012surface}.
Here, $d$ is the code distance indicating $2d^2 - 1$ physical qubits used per logical qubit, $p$ and $\varepsilon_d$ are the physical and logical error rates respectively, and $p_\text{thr}$ is the threshold error rate of the code.
Thus, when $p \ll p_\text{thr}$, the error rate of the logical qubit is suppressed exponentially in the distance of the code, with the error suppression factor $\Lambda = \varepsilon_{d}/\varepsilon_{d+2}  \approx p_\text{thr}/p$ representing the reduction in logical error rate when increasing the code distance by two.
While many platforms have demonstrated different features of quantum error correction~\cite{ryan2021realization, krinner2022realizing, sivak2023real, google2023suppressing, bluvstein2024logical, gupta2024encoding, da2024demonstration}, no quantum processor has definitively shown below-threshold performance.

Although achieving below-threshold physical error rates is itself a formidable challenge, fault-tolerant quantum computing also imposes requirements beyond raw performance.
These include features like stability over the hours-long timescales of quantum algorithms~\cite{gidney2021factor} and the active removal of correlated error sources like leakage~\cite{terhal2005fault}.
Fault-tolerant quantum computing also imposes requirements on classical coprocessors -- namely, the syndrome information produced
by the quantum device must be decoded as fast as it is generated~\cite{terhal2015quantum}.
The fast operation times of superconducting qubits, ranging
from tens to hundreds of nanoseconds, provide an advantage in speed but also a challenge for decoding errors
both quickly and accurately.

In this work, we realize surface codes operating below threshold on two superconducting processors.
Using a 72-qubit processor, we implement a distance-5 surface code operating with an integrated real-time decoder. 
In addition, using a 105-qubit processor with similar performance, we realize a distance-7 surface code.
These processors demonstrate $\Lambda > 2$ up to distance-5 and distance-7, respectively.
Our distance-5 quantum memories are beyond break-even, with distance-7 preserving quantum information for more than twice as long as its best constituent physical qubit.
To identify possible logical error floors, we also implement high-distance repetition codes on the 72-qubit processor, with error rates that are dominated by correlated error events occurring once an hour.
These errors, whose origins are not yet understood, set a current error floor of $10^{-10}$.
Finally, we show that we can maintain below-threshold operation on the 72-qubit processor even when decoding in real time, meeting the strict timing requirements imposed by the processor's fast 1.1\,\textmu s cycle duration.

\section{A Surface code memory below threshold}

\begin{figure*}[!ht]
    \centering
    \includegraphics[width=7.1in]{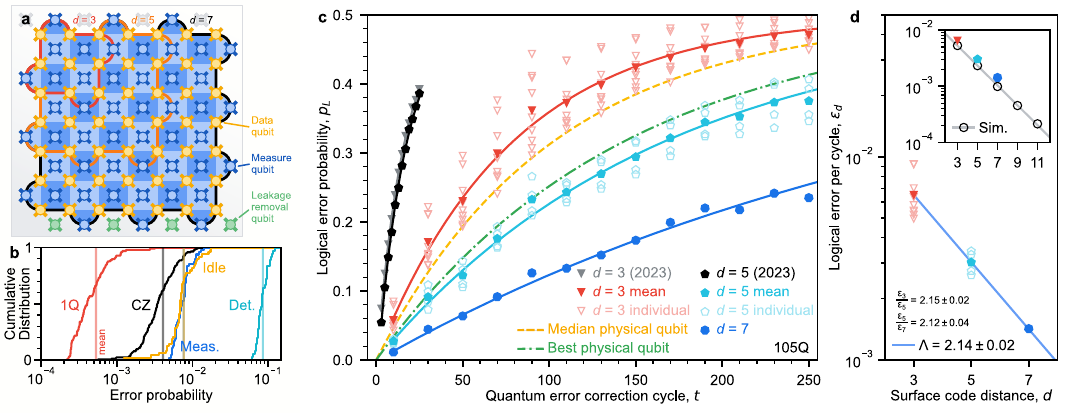}
    \caption{
    \textbf{Surface code performance.}
    \textbf{a,} Schematic of a distance-7 surface code on a 105-qubit processor.
    Each measure qubit (blue) is associated with a stabilizer (blue colored tile).
    Red outline: one of nine distance-3 codes measured for comparison ($3\times 3$ array).
    Orange outline: one of four distance-5 codes measured for comparison (4 corners).
    Black outline: distance-7 code.
    We remove leakage from each data qubit (gold) via a neighboring qubit below it, using additional leakage removal qubits at the boundary (green).
    \textbf{b,} Cumulative distributions of error probabilities measured on the 105-qubit processor.
    Red: Pauli errors for single-qubit gates.
    Black: Pauli errors for CZ gates.
    Blue: Average identification error for measurement.
    Gold: Pauli errors for data qubit idle during measurement and reset.
    Teal: weight-4 detection probabilities (distance-7, averaged over 250 cycles).
    \textbf{c,} Logical error probability, $p_L$, for a range of memory experiment durations.
    Each datapoint represents $10^5$ repetitions decoded with the neural network and is averaged over logical basis ($X_L$ and $Z_L$).
    Black and grey: data from Ref.~\cite{google2023suppressing} for comparison.
    Curves: exponential fits after averaging $p_L$ over code and basis. To compute $\varepsilon_d$ values, we fit each individual code and basis separately~\cite{supplement}.
    \textbf{d,} Logical error per cycle, $\varepsilon_d$, reducing with surface code distance, $d$.
    Uncertainty on each point is less than $5\times10^{-5}$.
    Symbols match panel c. 
    Means for $d=3$ and $d=5$ are computed from the separate $\varepsilon_d$ fits for each code and basis.
    Line: fit to Eq.~1, determining $\Lambda$.
    Inset: simulations up to $d=11$ alongside experimental points, both decoded with ensembled matching synthesis for comparison. Line: fit to simulation, $\Lambda_\textrm{sim}=2.25\pm 0.02$.
    }
    \label{surface_code}
\end{figure*}

We begin with results from our 105-qubit processor depicted in Fig.~\ref{surface_code}a.
It features a square grid of superconducting transmon qubits~\cite{koch2007charge} with improved operational fidelities compared to our previously reported processors~\cite{arute2019quantum, google2023suppressing}.
The qubits have a mean operating $T_{1}$ of 68\,\textmu s and $T_{2,\text{CPMG}}$ of 89\,\textmu s, which we attribute to improved fabrication techniques,  participation ratio engineering, and circuit parameter optimization.
Increasing coherence contributes to the fidelity of all of our operations which are displayed in Fig.~\ref{surface_code}b.

We also make several improvements to decoding, employing two types of offline high-accuracy decoders.
One is a neural network decoder~\cite{bausch2023learning}, and the other is a harmonized ensemble~\cite{shutty2024efficient} of correlated minimum-weight perfect matching decoders~\cite{fowler2013optimal} augmented with matching synthesis~\cite{jones2024improved}. 
These run on different classical hardware, offering two potential paths towards real-time decoding with higher accuracy.
To adapt to device noise, we fine-tune the neural network with processor data~\cite{bausch2023learning} and apply a reinforcement learning optimization to the matching graph weights~\cite{sivak2024optimization}.

We operate a distance-7 surface code memory comprising 49 data qubits, 48 measure qubits, and 4 additional leakage removal qubits, following the methods in Ref.~\cite{google2023suppressing}.
Summarizing, we initiate surface code operation by preparing the data qubits in a product state in either the $X_L$ or $Z_L$ basis of the \textit{ZXXZ} surface code~\cite{bonilla2021xzzx}.
We then repeat a variable number of cycles of error correction, during which measure qubits extract parity information from the data qubits to be sent to the decoder.
Following each syndrome extraction, we run data qubit leakage removal (DQLR)~\cite{miao2023overcoming} to ensure that leakage to higher states is short-lived.
We measure the state of the logical qubit by measuring the individual data qubits and then check whether the decoder's corrected logical measurement outcome agrees with the initial logical state.

From surface code data, we can characterize the physical error rate of the processor using the bulk error detection probability~\cite{hesner2024using}.
This is the proportion of weight-4 stabilizer measurement comparisons that detect an error.
The surface code detection probabilities are $p_\text{det}=(7.7\%, 8.5\%, 8.7\%)$ for $d=(3, 5, 7)$.
We attribute the increase in detection probability with code size to finite size effects~\cite{supplement} and parasitic couplings between qubits. 
We expect both effects to saturate at larger processor sizes~\cite{klimov2024optimizing}.

We characterize our surface code logical performance by fitting the logical error per cycle $\varepsilon_d$ up to 250 cycles, averaged over the $X$ and $Z$ bases.
We average the performance of 9 different distance-3 subgrids and 4 different distance-5 subgrids to compare to the distance-7 code.
Finally, we compute the error suppression factor $\Lambda$ using linear regression of $\ln(\varepsilon_d)$ versus $d$.
With our neural network decoder, we observe 
$\Lambda = 2.14 \pm 0.02$ and $\varepsilon_7$ = $(1.43 \pm 0.03)\times 10^{-3}$ 
(see Fig.~\ref{surface_code}c-d).
With ensembled matching synthesis, we observe
$\Lambda = 2.04 \pm 0.02$ and $\varepsilon_7$ = $(1.71 \pm 0.03)\times 10^{-3}$.

Furthermore, we simulate logical qubits of higher distances using a noise model based on the measured component error rates in Fig.~\ref{surface_code}b, additionally including leakage and stray interactions between qubits~\cite{google2023suppressing, supplement}.
These simulations are shown alongside the experiment in the inset of Fig.~\ref{surface_code}d, both decoded with ensembled matching synthesis.
We observe reasonable agreement with experiment and decisive error suppression, affirming that the surface codes are operating below threshold.

Thus far, we have focused on the error suppression factor $\Lambda$, since below threshold performance guarantees that physical qubit lifetimes and operational fidelities can be surpassed with a sufficiently large logical qubit.
In fact, our distance-7 logical qubit already has more than double the lifetime of its constituent physical qubits.
While comparing physical and logical qubits is subtle owing
to their different noise processes, we plot a direct comparison between logical error rate and physical qubit error rate averaged over $X$ and $Z$ basis initializations in Fig.~\ref{surface_code}c.
To quantify qubit lifetime itself, we average uniformly over pure states using the metric proposed in Refs.~\cite{sivak2023real, supplement}.
The distance-7 logical qubit lifetime is $291\pm 6$ \textmu s, exceeding the lifetimes of all the constituent physical qubits (median $85\pm 7$\,\textmu s, best $119\pm 13$\,\textmu s) by a factor of $2.4 \pm 0.3$. 
Our logical memory beyond break-even extends previous results using bosonic codes~\cite{ofek2016extending, ni2023beating, sivak2023real} to multi-qubit codes, and it is a critical step toward logical operation break-even.

\section{Logical error sensitivity}

\begin{figure}
    \centering
    \includegraphics[width=\linewidth]{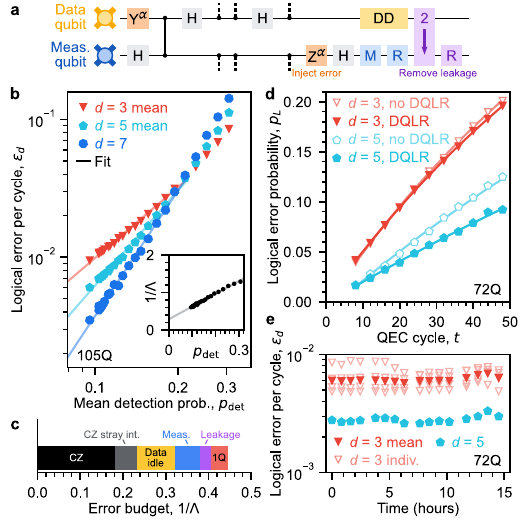}
    \caption{
    \textbf{Error sensitivity in the surface code.}
    \textbf{a,} One cycle of the surface code circuit, focusing on one data qubit and one measure qubit.
    Black bar: CZ, H: Hadamard, M: measure, R: reset, DD: dynamical decoupling.
    Orange: Injected coherent errors.
    Purple: Data qubit leakage removal (DQLR)~\cite{miao2023overcoming}.
    \textbf{b,} Error injection in the surface code.
    Distance-3 averages over 9 subset codes, and distance-5 averages over 4 subset codes, as in Fig.~\ref{surface_code}.
    Logical performance is plotted against the mean weight-4 detection probability averaging over all codes, where increasing the error injection angle $\alpha$ increases detection probability.
    Each experiment is 10 cycles with $2\times 10^4$ total repetitions.
    Lines: power law fits for data points at or below where the codes cross.
    Inset: Inverse error suppression factor, $1/\Lambda$, versus detection probability.
    Line: fit to points with $1/\Lambda < 1$.
    \textbf{c,} Estimated error budget for the surface code based on component errors and simulations.
    CZ: CZ error, excluding leakage and stray interactions.
    CZ stray int.: CZ error from unwanted interactions.
    Data idle: Data qubit idle error during measurement and reset.
    Meas.: Measurement and reset error.
    Leakage: Leakage during CZs and due to heating.
    1Q: Single-qubit gate error.
    \textbf{d,} Comparison of logical performance with and without data qubit leakage removal each cycle.
    Distance-3 points (red triangles) are averaged over four quadrants.
    Each experiment is $10^5$ repetitions.
    Curves: exponential fits.
    \textbf{e,} Repeating experiments to assess performance stability, comparing distance-3 and distance-5. Each point represents a sweep of logical performance versus experiment duration, up to 250 cycles.
    } \label{error_injection}
\end{figure}

Equipped with below-threshold logical qubits, we can now probe the sensitivity of logical error to various error mechanisms in this new regime. 
We start by testing how logical error scales with physical error and code distance. 
As shown in Fig.~\ref{error_injection}a, we inject coherent errors with variable strength on both data and measure qubits, and extract two quantities from each injection experiment. 
First, we use detection probability as a proxy for the total physical error rate.
Second, we infer the logical error per cycle by measuring logical error probability at 10 cycles, decoding with correlated matching~\cite{fowler2013optimal}.

In Fig.~\ref{error_injection}b, we plot logical error per cycle versus detection probability for the distance-3, -5, and -7 codes.
We find that the three curves cross near a detection probability of 20\%, roughly consistent with the crossover regime explored in Ref.~\cite{google2023suppressing}.
The inset further shows that detection probability acts as a good proxy for $1/\Lambda$.
When fitting power laws below the crossing, we observe approximately 80\% of the ideal value $(d$$+$$1)/2$ predicted by Eq.~1.
We hypothesize that this deviation is caused by excess correlations in the device.
Nevertheless, higher distance codes show faster reduction of logical error, realizing the characteristic threshold behavior \emph{in situ} on a quantum processor.

To quantify the impact of correlated errors along with more typical gate errors, we form an error budget.
Following the method outlined in Refs.~\cite{google2023suppressing, chen2021exponential}, we estimate the relative contribution of different component errors to $1/\Lambda$. 
We run simulations based on a detailed model of our 72-qubit processor.
The model includes local noise sources due to gates and measurements, as well as two sources of correlated error: leakage, and stray interactions during our CZ gates which can induce correlated $ZZ$ and swap-like errors~\cite{supplement}.
Fig.~\ref{error_injection}c shows our estimated error budget for $1/\Lambda$ in the 72-qubit processor when decoding with correlated matching.
Applying the same decoder to experimental data,
the error budget overpredicts $\Lambda$ by 20\%, indicating that most but not all error effects in our processor have been captured.
Correlated errors make up an estimated 17\% of the budget, and while not a dominant contributor, we expect their importance to increase as error rates decrease.
Overall, both local and correlated errors from CZ gates are the largest contributors to the error budget.
Consequently, continuing to improve both coherence and calibration will be crucial to further reducing logical error.

One potential source of excess correlations that we actively mitigate is leakage to higher excited states of our transmon qubits.
During logical qubit operation, we remove leakage from measure qubits using multi-level reset.
For data qubits, DQLR swaps leakage excitations to measure qubits (or additional leakage removal qubits)~\cite{miao2023overcoming}.
To examine sensitivity to leakage, we measure logical error probability of distance-3 and distance-5 codes in our 72-qubit processor with and without DQLR, with the results shown in Fig.~2d.
While activating DQLR does not strongly affect distance-3 performance, it substantially boosts distance-5 performance, resulting in a 35\% increase in $\Lambda$.
Comparatively, the detection probability decreases by only 12\%~\cite{supplement}, indicating that detection probability is only a good proxy for logical error suppression if errors are uncorrelated.
Overall, we find that addressing leakage is crucial to operating surface codes with transmon qubits~\cite{varbanov2020leakage, krinner2022realizing, miao2023overcoming}.

Finally, we test sensitivity to drift.
Using our 72-qubit processor, we measure logical performance of one distance-5 and four distance-3 codes 16 times over 15 hours, with the results shown in Fig.~\ref{error_injection}e.
Prior to the repeated runs, we employ a frequency optimization strategy which forecasts defect frequencies of two-level systems (TLS).
This helps to avoid qubits coupling to TLSs during the initial calibration as well as over the duration of the experiments.
Additionally, between every four experimental runs, we recalibrate the processor to account for potential qubit frequency and readout signal drift.
We observe an average $\Lambda = 2.18 \pm 0.07$ (standard deviation) and best $\Lambda = 2.31 \pm 0.02$~\cite{supplement} when decoding with the neural network.
While the performance of the worst distance-3 quadrant appears to fluctuate due to a transient TLS moving faster than our forecasts, this fluctuation is suppressed in the distance-5 code, suggesting that larger codes are less sensitive to component-level fluctuations.
Additionally, the logical error rates of experiments right after drift recalibration are not appreciably lower than those just prior, indicating that our logical qubit is robust to the levels of qubit frequency and readout drift present.
These results show that superconducting processors can remain stable over the hours-long timescales required for large scale fault-tolerant algorithms~\cite{gidney2021factor}.

\section{Probing the ultra-low error regime with repetition codes}

Despite realizing below-threshold surface codes, orders of magnitude remain between present logical error rates and the requirements for practical quantum computation.
In previous work running repetition codes, we found that high-energy impact events occurred approximately once every 10 seconds, causing large correlated error bursts which manifested a logical error floor around $10^{-6}$~\cite{google2023suppressing}.
Such errors would block our ability to run error-corrected algorithms in the future, motivating us to reassess repetition codes on our newer devices.

\begin{figure}
    \centering
    \includegraphics[width=\linewidth]{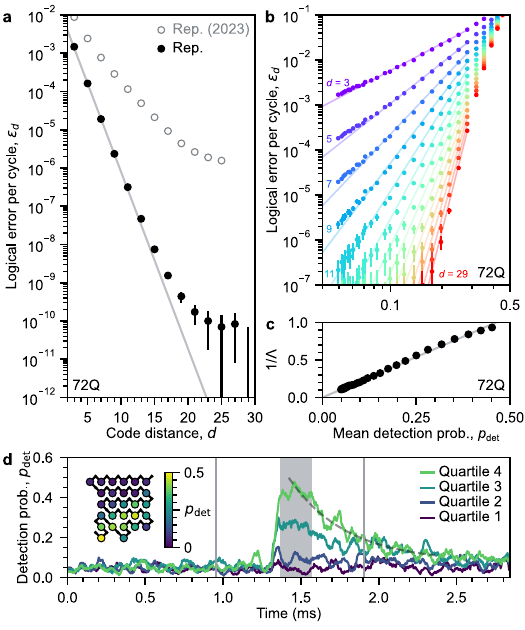}
    \caption{
    \textbf{High-distance error scaling in repetition codes.}
    \textbf{a,} Logical error per cycle, $\varepsilon_d$, versus code distance, $d$, when decoding with minimum-weight perfect matching.
    Repetition code points are from $d=29$, $10^3$-cycle experiments, $10^7$ repetitions for each basis $X$ and $Z$.
    We subsample smaller codes from the same $d=29$ dataset, averaging over subsamples.
    Line: fit of error suppression factor $\Lambda$.
    We include data from Rep.~\cite{google2023suppressing} for comparison.
    \textbf{b,} Logical error scaling with injected error.
    We inject a range of coherent errors on all qubits and plot against observed mean detection probability $p_{\text{det}}$.
    Each experiment is 10 cycles, and we average over $10^6$ repetitions.
    Smaller code distances are again subsampled from $d=29$.
    Lines: power law fits $\varepsilon_d = A_d p_\textrm{det}^{(d+1)/2}$ (one fit parameter, $A_d$), restricted to $\varepsilon_d > 10^{-7}$ and $p_\textrm{det} < 0.3$.
    \textbf{c,} $1/\Lambda$ scaling with injected error.
    Typical relative fit uncertainty is 2\%.
    Line: fit.
    \textbf{d,} Example event causing elevated detection probabilities which decay exponentially with time constant $369 \pm 6$\,\textmu s (gray dashed line).
    Three consecutive experimental shots are plotted, delimited by vertical gray lines.
    The 28 measure qubits are divided into four quartiles based on average detection probability in the gray-shaded window.
    Each trace represents the detection probability averaged over one quartile and a time window of 10 cycles.
    Inset: Average detection probability for each measure qubit (colored circle) within the gray-shaded window.
    } \label{repetition_code}
\end{figure}

Using our 72-qubit processor, we run a distance-29 repetition code for 1000 cycles of error correction over $2\times 10^7$ shots split evenly between bit- and phase-flip codes. 
In total, we execute $2 \times 10^{10}$ cycles of error correction comprising 5.5 hours of processor execution time.
Given the logical error probability $p_L$ at 1000 cycles, we infer the logical error per cycle as $\varepsilon_d = \frac{1}{2}\left(1 - (1 - 2p_L)^{1/1000}\right)$.
To assess how the logical error per cycle scales with distance-$d$, we follow Ref.~\cite{chen2021exponential} and subsample lower distance repetition codes from the distance-29 data.

Averaging over bit- and phase-flip repetition codes, we obtain an error suppression factor $\Lambda = 8.4 \pm 0.1$ when fitting logical error per cycle versus code distance between $d=5$ and 11, as shown in Fig.~\ref{repetition_code}a.
Notably, the error per cycle is suppressed far below $10^{-6}$, breaking past the error floor observed previously.
We attribute the mitigation of high-energy impact failures to gap-engineered Josephson junctions~\cite{mcewen2024resisting}.
However, at code distances $d\geq 15$, we observe a deviation from exponential error suppression at high distances culminating in an apparent logical error floor of $10^{-10}$.
Although we do not observe any errors at distance-29, this is likely due to randomly decoding correctly on the few most damaging error bursts.
While this logical error per cycle might permit certain fault-tolerant applications~\cite{campbell2021early}, it is still many orders of magnitude higher than expected and precludes larger fault-tolerant circuits~\cite{gidney2021factor, kivlichan2020improved}.

When we examine the detection patterns for these high-distance logical failures, we observe two different failure modes.
The first failure mode manifests as one or two detectors suddenly increasing in detection probability by over a factor of 3, settling to their initial detection probability tens or hundreds of cycles later~\cite{supplement}.
These less damaging failures could be caused by transient TLS's appearing near the operation frequencies of a qubit, or by coupler excitations, but might be mitigated using methods similar to Refs.~\cite{varbanov2020leakage, strikis2023quantum}.
The second and more catastrophic failure mode manifests as many detectors experiencing a larger spike in detection probability simultaneously; an example is shown in Fig.~\ref{repetition_code}d.
Notably, these anisotropic error bursts are spatially localized to neighborhoods of roughly 30 qubits (see inset).
Over the course of our $2\times 10^{10}$ cycles of error correction, our processor experienced six of these large error bursts, which are responsible for the highest-distance failures.
These bursts, such as the event shown in Fig.~\ref{repetition_code}d, are different from previously observed high-energy impact events~\cite{google2023suppressing}. 
They occur approximately once an hour, rather than once every few seconds, and they decay with an exponential time constant around 400\,\textmu s, rather than tens of milliseconds.
We do not yet understand the cause of these events, but mitigating them remains vital to building a fault-tolerant quantum computer.
These results reaffirm that long repetition codes are a crucial tool for discovering new error mechanisms in quantum processors at the logical noise floor.

Furthermore, while we have tested the scaling law in Eq.~1 at low distances, repetition codes allow us to scan to higher distances and lower logical errors.
Following a similar coherent error injection method as in the surface code, we show the scaling of logical error versus physical error and code distance in Fig.~\ref{repetition_code}b-c, observing good agreement with $O(p^{(d+1)/2})$ error suppression.
For example, reducing detection probability by a factor of 2 manifests in a factor of 250 reduction in logical error at distance-15, consistent with the expected $O(p^8)$ scaling.
This shows the dramatic error suppression that should eventually enable large scale fault-tolerant quantum computers, provided we can reach similar error suppression factors in surface codes.

\section{Real-time decoding}

In addition to a high-fidelity processor, fault-tolerant quantum computing also requires a classical coprocessor that can decode errors in real time.
This is because some logical operations are non-deterministic; they depend on logical measurement outcomes that must be correctly interpreted on the fly~\cite{terhal2015quantum}.
If the decoder cannot process measurements fast enough, an increasing backlog of syndrome information can cause an exponential blow-up in computation time.
Real-time decoding is particularly challenging for superconducting processors due to their speed.
The throughput of transmitting, processing, and decoding the syndrome information in each cycle must keep pace with the fast error correcting cycle time of 1.1\,\textmu s.
Using our 72-qubit processor as a platform, we demonstrate below-threshold performance alongside this vital module in the fault-tolerant quantum computing stack.

Our decoding system begins with our classical control electronics, where measurement signals are classified into bits then transmitted to a specialized workstation via low-latency Ethernet.
Inside the workstation, measurements are converted into detections and then streamed to the real-time decoding software via a shared memory buffer.
We employ a specialized version of the Sparse Blossom algorithm~\cite{higgott2023sparse} optimized to quickly resolve local configurations of errors common in surface code decoding, using a parallelization strategy similar to Ref.~\cite{wu2023fusion}.
The decoder operates on a constant-sized graph buffer which emulates the section of the error graph being decoded at any instant, but which does not grow with the total number of cycles used in the experiment.
Different threads are responsible for different spacetime regions of the graph,  processing their requisite syndrome information as it is streamed in~\cite{skoric2023parallel, tan2023scalable, wu2023fusion, bombin2023modular}.
These results are fused until a global minimum-weight perfect matching is found.
The streaming decoding algorithm is illustrated in Fig.~4a-b.
We also use a greedy edge reweighting strategy to increase accuracy by accounting for correlations induced by $Y$-type errors~\cite{fowler2013optimal, paler2023pipelined}.

\begin{figure}
    \centering
    \includegraphics[width=\linewidth]{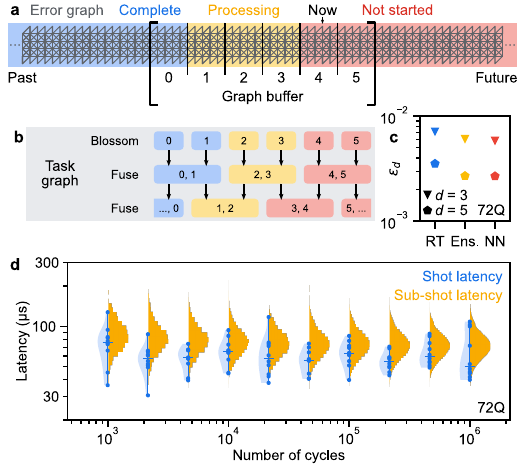}
    \caption{
    \textbf{Real-time decoding.}
    \textbf{a,} Schematic of the streaming decoding algorithm.
    Decoding problems are subdivided into blocks, with different threads responsible for different blocks.
    \textbf{b,} Task graph for processing blocks.
    Detections are allowed to match to block boundaries, which will then be processed downstream during a fuse step.
    If a configuration of detection events cannot be resolved by a future fuse step, the decoder heralds failure.
    We use 10-cycle blocks to ensure that the heralded failure rate is negligible compared to the logical failure rate.
    \textbf{c,} Accuracy comparison for the surface code with three decoders. We include the real-time decoder (RT), ensembled matching synthesis (Ens.), and the neural network decoder (NN). 
    Uncertainty on each point is less than $10^{-4}$~\cite{supplement}.
    \textbf{d,} Decoder latency versus experiment duration.
    The blue points correspond to end-of-shot latencies (10 shots per duration, horizontal bar: median, blue shading: violin plot).
    The yellow histograms represent sub-shot latencies obtained by checking how long after each 10-cycle block's data is received that the block is completed by the decoder.
    The sub-shot latencies tend to be slightly longer than end-of-shot latencies as the decoder may need to wait to fuse with detection events in future cycles in order to process up to the current cycle.
    }\label{decoding}
\end{figure}

In Fig.~\ref{decoding}d, we report the decoder latency, which we define as the time between the decoding software receiving the final cycle of syndrome measurements and the time when the decoder returns its correction.
For our distance-5 surface code, we test different problem sizes by increasing the number of error correction cycles up to $10^6$.
We observe that the average latency remains roughly constant at a net average of $63 \pm 17$\,\textmu s independent of the length of the experiment (up to 1.1\,seconds), indicating that the decoding problem is being processed in real time.
This latency will eventually determine the reaction time of the logical processor when enacting non-Clifford gates, but does not yet include feedback into the logical circuit.

Importantly, we are able to maintain below-threshold performance even under the strict timing requirements imposed by real-time decoding.
We run a dedicated experiment on our 72-qubit processor to compare real-time decoding to high-accuracy offline neural network decoding of the same data, with the results shown in Fig.~\ref{decoding}c.
Our real-time decoder achieves $\varepsilon_5 = 0.35\% \pm 0.01\%$ and $\Lambda = 2.0 \pm 0.1$ using a device-data-independent prior.
Meanwhile, the neural network decoder achieves $\varepsilon_5 = 0.269\% \pm 0.008\%$ and $\Lambda = 2.18 \pm 0.09$ when later decoding offline.
The modest reduction in accuracy when comparing the real-time decoder to an offline decoder is expected as the real-time decoder must operate significantly faster.
It requires a throughput of less than 1.1\,\textmu s per cycle compared to the neural network’s 24 \,\textmu s per cycle~\cite{bausch2023learning}.
However, we do expect that many of our high-accuracy decoding methods can eventually be made real-time by introducing techniques like layered or windowed decoding~\cite{skoric2023parallel, tan2023scalable, shutty2024efficient}.

\section{Outlook}

In this work, we have demonstrated a below-threshold surface code memory.
Each time the code distance increases by two, the logical error per cycle is reduced by more than half, culminating in a distance-7 logical lifetime more than double its best constituent physical qubit lifetime.
This signature of exponential logical error suppression with code distance forms the foundation of running large scale quantum algorithms with error correction.

Our error-corrected processors also demonstrate other key advances towards fault-tolerant quantum computing.
We achieve repeatable performance over several hours and run experiments up to $10^6$ cycles without deteriorating performance, both of which are necessary for future large scale fault-tolerant algorithms.
Furthermore, we have engineered a real-time decoding system with only a modest reduction in accuracy compared to our offline decoders.

Even so, many challenges remain ahead of us.
Although we might in principle achieve low logical error rates by scaling up our current processors, it would be resource intensive in practice.
Extrapolating the projections in Fig.~\ref{surface_code}d, achieving a $10^{-6}$ error rate would require a distance-27 logical qubit using 1457 physical qubits.
Scaling up will also bring additional challenges in real-time decoding as the syndrome measurements per cycle increase quadratically with code distance.
Our repetition code experiments also identify a noise floor at an error rate of $10^{-10}$ caused by correlated bursts of errors.
Identifying and mitigating this error mechanism will be integral to running larger quantum algorithms.

However, quantum error correction also provides us exponential leverage in reducing logical errors with processor improvements.
For example, reducing physical error rates by a factor of two would improve the distance-27 logical performance by four orders of magnitude, well into algorithmically-relevant error rates~\cite{campbell2021early, kivlichan2020improved}. 
We further expect these overheads will reduce with advances in error correction protocols~\cite{litinski2022active, chamberland2022universal, bravyi2024high, xu2024constant, gidney2023yoked, gidney2024inplace, cain2024correlated} and decoding~\cite{smith2023local,barber2023real,liyanage2024fpga}.

The purpose of quantum error correction is to enable large scale quantum algorithms. 
While this work focuses on building a robust memory, 
additional challenges will arise in logical computation~\cite{gidney2022stability, lin2024spatially}.
On the classical side, we must ensure that software elements including our calibration protocols, real-time decoders, and logical compilers can scale to the sizes and complexities needed to run multi-surface-code operations~\cite{bombin2023logical}.
With below-threshold surface codes, we have demonstrated processor performance that can scale in principle, but which we must now scale in practice.

\section{Author Contributions}
The Google Quantum AI team conceived and designed the experiment. The theory and experimental teams at Google Quantum AI developed the data analysis, modeling and metrological tools that enabled the experiment, built the system, performed the calibrations, and collected the data. All authors wrote and revised the manuscript and the Supplementary Information.
\vfill\null 
\pagebreak[4]
\section{Acknowledgements}
We are grateful to S.~Brin, S.~Pichai, R.~Porat, and J.~Manyika for their executive sponsorship of the Google Quantum AI team, and for their continued engagement and support.

\section{Ethics Declarations}
The authors declare no competing interests.

\section{Additional Information}
Supplementary Information is available for this paper.
Correspondence and requests for materials should be addressed to H. Neven (neven@google.com).

\section{Data Availability}
The data that support the findings of this study are available 
at~\url{https://doi.org/10.5281/zenodo.13273331}.

\clearpage
\onecolumngrid

\vspace{1em}
\begin{flushleft}
{\small Google Quantum AI and Collaborators}

\bigskip
{\small
\renewcommand{\author}[2]{#1$^\textrm{\scriptsize #2}$}
\renewcommand{\affiliation}[2]{$^\textrm{\scriptsize #1}$ #2 \\}

\newcommand{\xGoogle}{\affiliation{1}{Google Research}}

\newcommand{\xUMass}{\affiliation{2}{Department of Electrical and Computer Engineering, University of Massachusetts, Amherst, MA}}

\newcommand{\xGDM}{\affiliation{3}{Google DeepMind}}

\newcommand{\xUCSB}{\affiliation{4}{Department of Physics, University of California, Santa Barbara, CA}}

\newcommand{\xStorrs}{\affiliation{5}{Department of Physics, University of Connecticut, Storrs, CT}}

\newcommand{\xAuburnECE}{\affiliation{6}{Department of Electrical and Computer Engineering, Auburn University, Auburn, AL}}

\newcommand{\xETHZurich}{\affiliation{7}{Department of Physics, ETH Zurich, Switzerland}}

\newcommand{\xMITLab}{\affiliation{8}{Research Laboratory of Electronics, Massachusetts Institute of Technology, Cambridge, MA}}

\newcommand{\xMITEE}{\affiliation{9}{Department of Electrical Engineering and Computer Science, Massachusetts Institute of Technology, Cambridge, MA}}

\newcommand{\xMITPhysics}{\affiliation{10}{Department of Physics, Massachusetts Institute of Technology, Cambridge, MA}}

\newcommand{\xUCRPA}{\affiliation{11}{Department of Physics and Astronomy, University of California, Riverside, CA}}

\newcommand{\xYaleQI}{\affiliation{12}{Yale Quantum Institute, Yale University, New Haven, CT}}

\newcommand{\xYalePhysics}{\affiliation{13}{Departments of Applied Physics and Physics, Yale University, New Haven, CT}}

\newcommand{\Google}{1}
\newcommand{\UMass}{2}
\newcommand{\GDM}{3}
\newcommand{\UCSB}{4}
\newcommand{\Storrs}{5}
\newcommand{\AuburnECE}{6}
\newcommand{\ETHZurich}{7}
\newcommand{\MITLab}{8}
\newcommand{\MITEE}{9}
\newcommand{\MITPhysics}{10}
\newcommand{\UCRPA}{11}
\newcommand{\YaleQI}{12}
\newcommand{\YalePhysics}{13}

\author{Rajeev Acharya}{\Google},
\author{Laleh \ Aghababaie-Beni}{\Google},
\author{Igor Aleiner}{\Google},
\author{Trond I.~Andersen}{\Google},
\author{Markus Ansmann}{\Google},
\author{Frank Arute}{\Google},
\author{Kunal Arya}{\Google},
\author{Abraham Asfaw}{\Google},
\author{Nikita Astrakhantsev}{\Google},
\author{Juan Atalaya}{\Google},
% ...

\bigskip

\xGoogle
\xUMass
\xGDM
\xUCSB
\xStorrs
\xAuburnECE
\xETHZurich
\xMITLab
\xMITEE
\xMITPhysics
\xUCRPA
\xYaleQI
\xYalePhysics

}
\end{flushleft}

\clearpage
\twocolumngrid
\bibliographystyle{naturemag}
\bibliography{bibliography}

\end{document}


\preprint{APS/123-QED}

\title{Supplementary Information for \\``Quantum error correction below the surface code threshold''}

\author{Google Quantum AI and Collaborators} \noaffiliation

\date{\today}

\maketitle

\renewcommand{\thetable}{S\arabic{table}}  
\renewcommand{\thefigure}{S\arabic{figure}}
\renewcommand{\arraystretch}{1.25}

\tableofcontents
\section{Gates and Readout}
\input{text_sm/device_characterization}

\section{Improvements in Control Techniques}
\subsection{Frequency Optimization}
\input{text_sm/sys_opt}

\section{Decoding}
\input{text_sm/decoder_intro}
\subsection{Decoder Priors}
\input{text_sm/decode_priors}
\subsection{Neural Network Decoder Details}
\input{text_sm/ML_decoder_details}
\subsection{Real-Time Decoding System }
\input{text_sm/real_time_decoding}

\section{Understanding Fidelity}
\subsection{Error Budget Simulations}
\input{text_sm/error_budget_sims}

\subsection{Comparison to Physical Qubit Lifetime}
\input{text_sm/break_even}

\section{Error Correction Experimental Details}
\subsection{Low Probability Events in the Repetition Code}
\input{text_sm/rep_code_floor}
\subsection{Grid and Circuit Details}
\input{text_sm/qecx_details}
~
\section{Uncertainty Analysis}
\input{text_sm/uncertainty}
\clearpage
\bibliographystyle{naturemag}
\bibliography{references_sm}

%% file: text_sm/device_characterization.tex
Single-qubit gates are implemented using microwave $XY$ rotations and virtual $Z$ rotations. 
On the 105-qubit processor, all $XY$ rotations have 25\,ns duration, while on the 72-qubit processor, $\pi$ and $\pi/2$ rotations have 35\,ns duration and 18\,ns duration respectively.
We implement our entangling operations using CZ gates. 
Our CZ gates are performed in 42\,ns on the 105-qubit processor, and 37\,ns on the 72-qubit processor. 

Our measurement chain is similar to that of \cite{google2023suppressing}, though with Lumped Element Snake Amplifiers (LESAs) \cite{PhysRevApplied.20.054058} replacing the previous impedance matched parametric amplifiers.
The qubit frequency during readout, readout pulse duration, and readout pulse power are jointly optimized \cite{PhysRevLett.132.100603} to attain simultaneous high fidelity mid-circuit readout. 

\begin{figure}
    \includegraphics{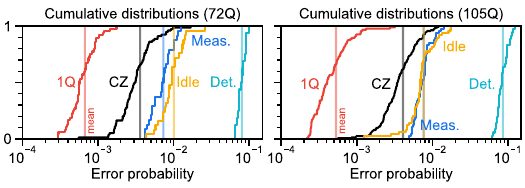}
    \caption{\textbf{Operation error probabilities.}
        See Fig.\,~1b.
        Left: distribution of benchmark results for the 72-qubit processor configured to run the distance-5 surface code.
        Right: 105-qubit processor configured to run the distance-7 surface code (same as Fig.\,~1b).
    }
    \label{fig:edcfs}
\end{figure}

In main text Fig.\,~1b, we show distributions of operation error probabilities for the 105-qubit processor. 
In Fig.\,~\ref{fig:edcfs}, we repeat Fig.\,~1b alongside equivalent measurements on the 72-qubit processor.

%% file: text_sm/sys_opt.tex
We optimize single- and two-qubit gate frequency trajectories via the Snake optimizer \cite{snake2020}. We leverage technology developed in Ref. \cite{snake2024}, highlighting two strategies that were especially helpful in boosting data qubit coherence. First, we constructed our optimization model directly around the error correction circuit. This strategy prioritizes data qubit coherence to a much greater extent than optimizing for interleaved layers of single- and two-qubit gates as in cross-entropy benchmarking. Second, we employed intermediate-dimensional optimization up to 9 dimensions. This strategy enables more complex trade-offs between data and measure qubits than lower dimensional optimization, without significantly compromising runtime.

To ensure that qubit coherence remained stable over multiple days of datataking, two stability optimization strategies were included into the Snake optimizer. Both were intended to mitigate catastrophic losses in coherence from frequency collisions with TLS, which are known to have time-dependent variations in their transition frequencies \cite{Klimov:2018nem}. Both strategies leveraged information extracted from spectrally and temporally resolved qubit $T_{1}$ data. The first strategy modified the relaxation error component to penalize single-qubit gate frequencies which historically had reduced $T_{1}$. The second strategy sought to prevent frequency collisions with TLS by extracting frequency-temporal trajectories of TLS from historical $T_{1}$ data, forecasting them to future times, and excluding these frequencies from allowed single-qubit gate frequencies.

%% file: text_sm/decoder_intro.tex
\subsection{List of Decoders}

\begin{table*}[htb!]
\centering
    \begin{tabular}{| c|
     >{\centering} m{10em}
     >{\centering} m{10em}|
     >{\centering} m{10em}
     >{\centering} m{10em}
     c|}
    \hline
      {} & \multicolumn{2}{c|}{\textbf{Fig.~1d}}   & \multicolumn{3}{c|}{\textbf{Fig.~4c}} \\ [0.5ex]  
         & Libra & Neural network & Real-time & Libra & Neural network \\ \hline
        \hspace{1em}$\varepsilon_3$\hspace{1em} & $(7.12 \pm 0.06)\times 10^{-3}$ & $(6.50 \pm 0.06)\times 10^{-3}$ & $(7.1 \pm 0.3)\times 10^{-3}$ & $(6.1 \pm 0.2)\times 10^{-3}$ & $(5.9 \pm 0.1)\times 10^{-3}$ \\[0.5ex]
        $\varepsilon_5$ & $(3.49 \pm 0.04)\times 10^{-3}$ & $(3.03 \pm 0.03)\times 10^{-3}$ & $(3.5 \pm 0.1)\times 10^{-3}$ & $(2.70 \pm 0.08)\times 10^{-3}$ & $(2.69 \pm 0.08)\times 10^{-3}$ \\[0.5ex]
        $\varepsilon_7$ & $(1.71 \pm 0.03)\times 10^{-3}$ & $(1.43 \pm 0.03)\times 10^{-3}$ &  &  &  \\
    \hline
    \end{tabular}
    \caption{\textbf{Logical errors per cycle for different decoding schemes.} Here we report the logical error per cycle, $\varepsilon_d$, for different code distances $d$ and different decoding schemes (real-time, Libra~\cite{jones2024improved}, and neural network~\cite{bausch2023learning}). These correspond to the values plotted in Fig.~1d and Fig.~4c of the main text.} 
    \label{Suppl:table-logical-error-rates}
\end{table*}

In this work, we use several different decoders depending on the context of the experiment.
We use, 
\begin{enumerate}
\item a neural network decoder introduced in Ref.~\cite{bausch2023learning},
\item a matching synthesis decoder Libra recently introduced in Ref.~\cite{jones2024improved} with an ensemble size of 51,
\item an ensembled matching decoder Harmony introduced in Ref.~\cite{shutty2024efficient} with an ensemble size of 101 (unless otherwise specified),
\item a correlated matching decoder similar to that described in Ref.~\cite{fowler2013optimalcomplexitycorrectioncorrelated},
\item a real-time decoder with accuracy similar to Ref.~\cite{Paler2023pipelinedcorrelated}. 
\end{enumerate}

Predictions for these decoders applied to various datasets reported in the paper, along with additional datasets not reported here, can be found at Ref.~\cite{qec_datasets_zenodo_2024}.
We note there are other accurate decoders that would be interesting to test on this data, including beliefmatching~\cite{higgott2023improved}, BP+OSD~\cite{roffe_decoding_2020}, and tensor network decoding~\cite{google2023suppressing}.

%% file: text_sm/decode_priors.tex
We use two methods for configuring decoder priors. In experiments where maximizing the performance is not the main objective (error injection, real-time decoding demo), we use the SI1000 error model \cite{gidney2021fault}. This error model is inspired by the typical hierarchy of error rates in  superconducting qubits, but is otherwise agnostic to the device and decoder.

In experiments that require the highest accuracy, we use the learning-based approach for calibrating the decoder priors \cite{sivak2024optimization}.
In the learning-based approach, we adopt the hyperparameters from Ref.~\cite{sivak2024optimization} for the surface code and train the prior on a ``calibration dataset" of duration 13 cycles, which precedes the rest of the experiment. The learned parameters of the prior are then used for decoding all datasets up to 250 cycles. Although we frequently decode the surface code using ensembles of matching decoders~\cite{shutty2024efficient, jones2024improved}, we use a faster correlated-matching decoder~\cite{higgott2023sparse} for training. As was shown in \cite{sivak2024optimization}, the learned priors generalize well across these two decoders. 

Our optimization of the priors relies on the sensor technique detailed in Ref.~\cite{sivak2024optimization}. For the distance-7 surface code, smaller patches of distance-3 and distance-5, shown in Fig.~\ref{fig:sub_grids_supplement}(a,b), act as local sensors of the error landscape in the device. These sensor-codes are used to optimize the parameters of the prior, which is then applied to decoding the target distance-7 code. The sensor technique is especially important in the repetition code, as using the logical error rate (LER) of the target distance-29 code in the optimization loop becomes computationally infeasible. Instead, we optimize the prior with the help of 25 distance-5 sensors subsampled from the target code. For the minimum-weight perfect matching (MWPM) decoder, we find that training the prior results in $10\%$ average improvement of the LER at all distances relative to the SI1000 prior. 

\begin{figure*}
  \centering
  \includegraphics{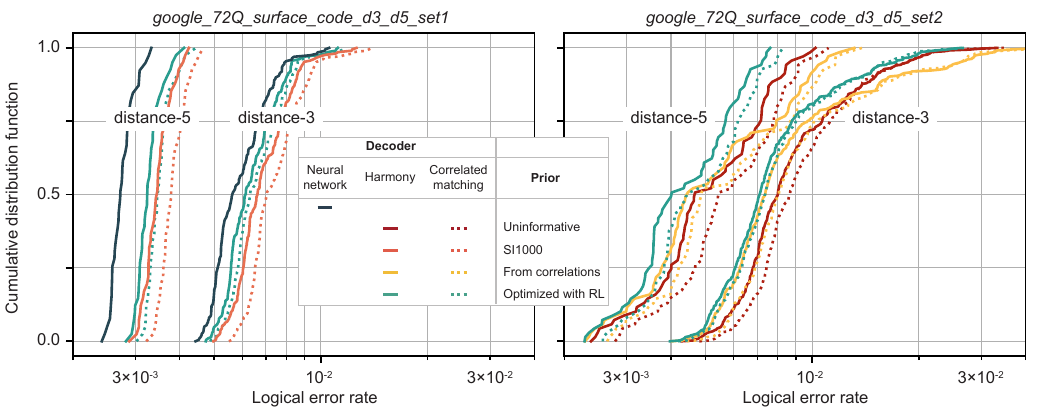}
  \caption{\textbf{Characterization of offline decoding methods.} The decoder calibration methods benchmarked here are the same as in Ref.~\cite{sivak2024optimization}. The first dataset corresponds to the 15-hour sweep in Fig.~2(e) of the main text. The second dataset contains experiments in which the device was well calibrated and experiments in which the calibrations have not been refreshed for several days. This selection is intended for testing the decoding algorithms under a wide range of experimental conditions.}
  \label{fig:super_weird_figure}
\end{figure*}

Additional characterization of decoders and priors on two datasets, \texttt{google\_72Q\_surface\_code\_d3\_d5\_set1} and \texttt{google\_72Q\_surface\_code\_d3\_d5\_set2}, is shown in Fig.~\ref{fig:super_weird_figure}. The summary of the datasets, which we made publicly available in Ref.~\cite{qec_datasets_zenodo_2024}, is presented in Table~\ref{Suppl:table-released-datasets}.

On the first dataset, we find that the neural network decoder \cite{bausch2023learning} achieves the highest accuracy overall. Among the matching-based decoders we ran for Table~\ref{Suppl:table-released-datasets}, the most performant configuration is Harmony \cite{shutty2024efficient},
although we expect that augmenting with matching synthesis would further boost accuracy \cite{jones2024improved}. 
We also use a reinforcement learning optimized prior \cite{sivak2024optimization}. On the distance-5 code, just ensembling leads to a $1.14$ times lower LER on average than correlated matching with SI1000 prior. 
We expect this improvement will apply to real-time experiments as well, once ensembled matching is layered, parallelized, and incorporated into the real-time decoder, and if optimization of the prior can be done on a time scale faster than the device drift~\cite{sivak2024optimization}. We consider both of these requirements feasible.
We also note that while this characterization uses Harmony as described in \cite{shutty2024efficient}, the main text uses the recently-developed ensembled matching-synthesis decoder Libra~\cite{jones2024improved}.

\begin{table*}
\centering
    \begin{tabular}{|c|c|c|c|c|c|}
    \hline
      {{\bf Dataset}} & \begin{tabular}{c} {\bf Number of} \\ {\bf experiments} \end{tabular} & {\bf Basis} & {\bf Distance} & {\bf Cycles} & {\bf Shots} \\ \hline
      \begin{tabular}{p{8.5cm}}
      \texttt{google\_72Q\_surface\_code\_d3\_d5\_set1} \\
      This surface code dataset corresponds to Fig. 2(e) in the main text, which represents the last 16 experiments that were run consecutively.
      \end{tabular}
      & 21 & $X$, $Z$ & 3, 5 & $1$ to $250$ & $5\times10^4$          
      \\ \hline
      \begin{tabular}{p{8.5cm}}
      \texttt{google\_72Q\_surface\_code\_d3\_d5\_set2}  \\
      Samples in this surface code dataset capture different calibration states of the device, ranging from well-calibrated with LER of the distance-5 code smaller than $3\times 10^{-3}$ to poorly calibrated with LER approaching $10^{-2}$. This selection is intended for testing the decoding algorithms under a wide range of experimental conditions.
      \end{tabular}
      & 35 & $X$, $Z$ & 3, 5 & $5$ to $50$ & $6\times10^4$          
      \\ \hline
      \begin{tabular}{p{8.5cm}}
      \texttt{google\_105Q\_surface\_code\_d3\_d5\_d7}  \\
      This surface code dataset corresponds to Fig. 1(c,d) in the main text.
      \end{tabular}
      & 1 & $X$, $Z$ & 3, 5, 7 & $1$ to $250$ & $5\times10^4$ 
      \\ \hline
      \begin{tabular}{p{8.5cm}}
      \texttt{google\_72Q\_repetition\_code\_d29}  \\
      This repetition code dataset corresponds to Fig. 3(a) in the main text.
      \end{tabular}
      & 100 & $X$, $Z$ & 29 & $10^3$ & $10^5$ 
      \\ \hline
    \end{tabular}
    \caption{\textbf{QEC datasets.} Brief summary of the QEC datasets released in Ref.~\cite{qec_datasets_zenodo_2024}. Further details can be found in each dataset's \texttt{README.md} file, including examples of data processing with open-source tools such as Stim \cite{gidney2021stim} and PyMatching \cite{higgott2022pymatching}.}
    \label{Suppl:table-released-datasets}
\end{table*}

%% file: text_sm/ML_decoder_details.tex
The machine learning decoder is the recurrent attention-based neural network described in \cite{bausch2023learning} which is trained to predict the logical error based on the stabilizer measurements of a surface code experiment. The network processes one cycle of stabilizer measurements at a time, updating an internal state representation which consists of a vector for each stabilizer. At the end of the experiment the state vectors are combined to produce a calibrated error probability.

As before, the networks (for both the 72-qubit and 105-qubit processors) are trained in two stages. We first pretrain on synthetic data from a generic noise model (SI1000 at $p=0.4\%$ in this case) generated by Stim \cite{gidney2021stim}. Separate models are trained for $X$ and $Z$ bases for distance-3, distance-5 and distance-7 experiments. We trained 5 separate models with different random seeds for each condition, resulting in 10 pretrained models per code distance that were the basis for all subsequent fine-tuning.

After pretraining, each network is further trained to match a specific experiment by fine-tuning using a limited quantity of experimental data. This fine-tuning stage is broken up into two steps: a first step during which we fine-tune with samples from a $p_{ij}$ detector error model fitted to the 13 cycle dataset, taken before the actual experiment \cite{google2023suppressing}. In a second step the model is fine-tuned on the experimental samples themselves.

For fine-tuning on experimental data, and as in previous experiments \cite{bausch2023learning}, we split the experimental data into multiple partitions for cross-validation. We further subdivide each partition into a training and validation set to choose an early-stopping time to prevent overfitting. Test performance is measured on the held out partition. Hyperparameters are chosen based on previous datasets, so no other evaluations are performed on this data.

We utilize various splits of the data. For the synthetic dataset, of which there are two identical copies, we fine-tune on one and test on the other. For the realtime and distance-7 datasets, of which there is only one, we utilize an even/odd split. For the “stability” experiment with multiple temporally spaced repeated memory experiments, we chose a truly causal scenario: For the $N$\textsuperscript{th} dataset, we fine-tune with the detector error model (DEM) data from the $N$\textsuperscript{th} set (since it was dependent only on the 13-cycle data captured at the start of that set) but subsequently fine-tune on the previous $(N-1)$\textsuperscript{th} dataset. We find that this strict temporal split leads to similar accuracy as an even/odd split.

We use the architecture from Ref.\,\cite{bausch2023learning} without attention bias, without the residual network before the recurrent neural network and with only two layers in the readout network.  Because the experimental data in this paper is of much longer duration (250 cycles vs.\ 25 for the data in \cite{google2023suppressing}) we add gated recurrence to the state update \cite{Cho2014-ne}. Instead of adding the state and incoming stabilizer representations, these are concatenated and projected through a 3-layer multilayer perceptron to compute element-wise update and reset gates. The reset-gated state and new embeddings are concatenated before projection and processing by the syndrome transformer layers. The final output and state are summed, weighted by the update gate. We find this made the training more stable for long-duration experiments. We mostly train the network on synthetic data of up to 25 cycles, but periodically sample examples of lengths up to 200 to ensure generalization to long experiments. The simulated SI1000 examples has logical observable labels for every cycle up to and including the total length.

\begin{table*}[htb!]
\centering
    \begin{tabular}{r
     >{\centering} m{4.5em}
     >{\centering} m{4.5em}
     >{\centering} m{4.5em}
     >{\centering} m{4.5em}
     >{\centering} m{4.5em} 
     >{\centering} m{4.5em}
     >{\centering} m{4.5em} 
     >{\centering} m{4.5em} 
     c}
    \hline
      {} & \multicolumn{3}{c}{\bf{Pretrain}}   & \multicolumn{3}{c}{\bf{Fine-tune} (DEM)} & \multicolumn{3}{c}{\bf{Fine-tune (Experimental)}}\\ [0.5ex]  
      & 3$\times$3 & 5$\times$5 & 7$\times$7 & 3$\times$3 & 5$\times$5 & 7$\times$7 & 3$\times$3 & 5$\times$5 & 7$\times$7 \\ \hline 
      \bf{Parameter count} & 5.411M & 5.419M & 5.432M & & & & &\\ [0.5ex]
      \bf{Channels} & 256 & 256 & 256 & & & & & & \\[0.5ex]
      \bf{Training steps} & 1.56M & 5.86M & 5.86M & 195k & 781k & 10.9M & 234k/15k$^*$ & 234k/15k$^*$ & 234k/15k$^*$ \\ [0.5ex]
      \bf{Batch size} & 256 & 256 & 256 & 256 & 256 & 256 & 64/1024$^*$ & 64/1024$^*$ & 64/1024$^*$ \\[0.5ex]
      \bf{Training examples} & 400M & 1500M & 1500M & 50M & 200M & 2800M & 15M$^\dagger$ & 15M$^\dagger$ & 15M$^\dagger$ \\[0.5ex]
      \bf{Learning rate} & 5e-6 & 5e-6 & 5e-6 & 1.7e-6 & 1.3e-6 & 2e-6 & 2e-7 & 2.5e-7 & 3e-7 \\[0.5ex]
      \bf{Weight decay} & 1e-5 & 1e-5 & 1e-5 & 1e-5 & 1e-4 & 1e-3 & 7e-5 & 7e-5 & 7e-5 \\[0.5ex]
      \bf{Cosine cycle} & 250M & 300M & 500M & 40M & 160M & 640M & 30M & 80M & 200M \\[0.5ex]
      \bf{EMA step} & 1e-4 & 1e-4 & 1e-4 & 1e-4 & 1e-4 & 1e-4 & 8e-4 & 8e-4 & 8e-4 \\ [0.5ex]
    \hline
    \end{tabular}
    \caption{\textbf{Hyperparameters for the machine learning decoder.} $^*$For fine-tuning on the realtime and simulated data we choose a batch size of 1024; for the others a batch size of 64. $^\dagger$For fine-tuning with the even-odd (causal) split, we train on 258,440 (583,440) unique experimental samples: we keep 5120 validation set samples of each of the 25,000 (50,000) shots per split for each of the 13 experiment durations (10, 30, …, 250 cycles). For the real-time experiment we only have 5,000 examples per duration, so we train on 1700 and tested on 800 from each split. Consequently, training presents each example many times. For pretraining and DEM fine-tuning we are sampling data from a simulator so each training example is different.} 
    \label{Suppl:ML_decoding_table}
\end{table*}

%% file: text_sm/real_time_decoding.tex
\subsubsection{Technical Requirements}
In this section we provide a brief overview of general technical requirements for a real-time syndrome decoding system and describe the key aspects of the specific system employed in this work.

In the simplest quantum error correction experiments, syndrome data is saved in a file and processed by a decoder offline after the quantum circuit is finished executing on the quantum processor. This offline mode of syndrome decoding is insufficient for experiments involving fault-tolerant constructions, such as gate teleportation \cite{Gottesman1999} that use feed-forward, i.e. the capability to condition quantum operations on measurement outcomes. In order to support these more advanced experiments, one needs a fast decoder capable of computing logical measurement outcomes in real-time during the execution of a quantum circuit.

A real-time syndrome decoding system must satisfy requirements concerning three key performance metrics: accuracy, latency, and throughput. Decoding accuracy is characterized by the per-cycle probability that the logical measurement outcome computed by the decoder is correct.

Decoding latency, $T_\text{decode}$, is the delay between when the measurement outcomes of the physical qubits comprising a logical qubit become known and when the measurement outcome of the logical qubit becomes known. This is an important contributor to the overall reaction time, $T_\text{react}$, which is the time it takes the quantum computer's control system to react to a logical measurement by executing a dependent logical operation \cite{gidney2019flexiblelayoutsurfacecode}. Inverse reaction time is the rate at which quantum computer can execute logical operations that require feed-forward for fault-tolerance, such as the $T$-gate in the surface code. Thus, inverse reaction time plays the role of clock speed of a quantum computer.

In more detail, the reaction time is given by the sum
\begin{align}
    T_\text{react} = T_\text{decode} + T_\text{control}
\end{align}
of the decoding latency $T_\text{decode}$ and the time $T_\text{control}$ that the control system spends executing quantum gates, obtaining measurement outcomes for physical qubits, and reacting to the logical measurement outcomes provided by the real-time decoding system. In a non-error-corrected processor with feed-forward, we have $T_\text{decode}=0$ and the reaction time is just $T_\text{control}$. Decoding latency $T_\text{decode}$ constitutes most of the overhead of the quantum error correction layer.

In a software-based real-time syndrome decoding system, we can break decoding latency down into software latency and I/O latencies,
\begin{align}
    T_\text{decode} = T_\text{input} + T_\text{software} + T_\text{output}
\end{align}
where $T_\text{software}$ is the latency of the software components of the decoding system, and $T_\text{input}$ and $T_\text{output}$ are the input and output latencies of the hardware-software interface which facilitates exchange of information between the quantum computer's control electronics and the decoding software.

Another requirement for a real-time decoding system concerns its throughput, i.e. the amount of syndrome information it is capable of processing per unit time. Decoding throughput must be at least as high as the amount of syndrome information generated per unit time. Failure to satisfy this requirement causes syndrome backlog to accumulate and leads to exponential slowdown of the quantum computer \cite{RevModPhys.87.307}.

\begin{figure*}
  \centering
  \includegraphics[width=1.95\columnwidth]{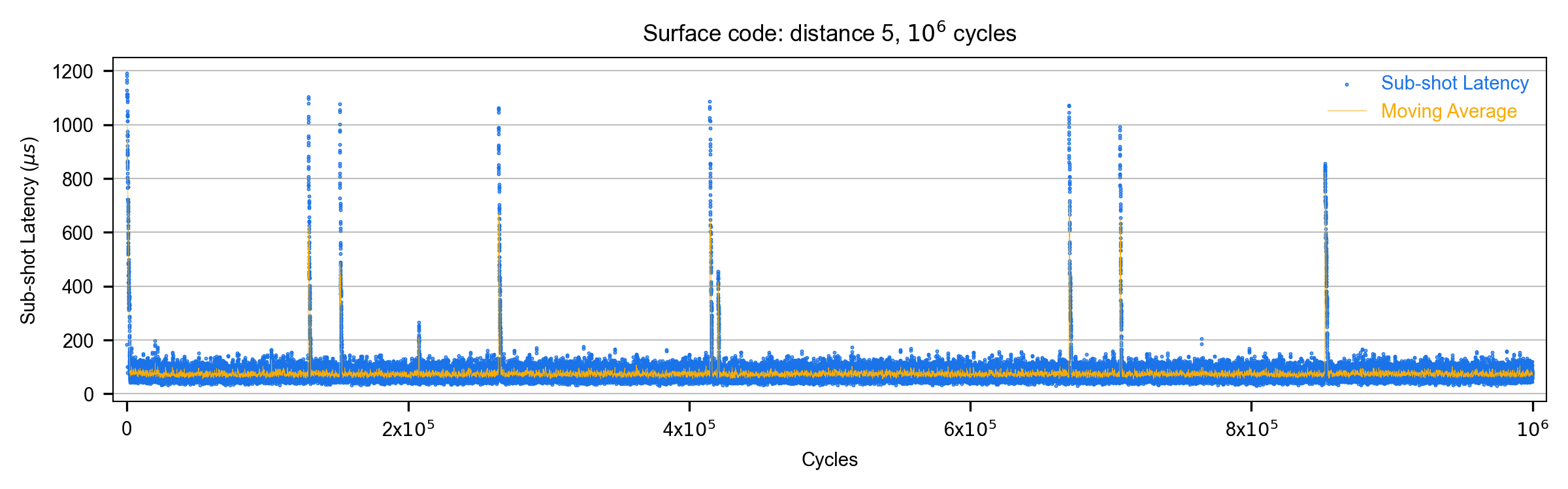}
  \caption{\textbf{Sub-shot latency as a function of cycles within a single experiment.} Sub-shot latency is defined as the delay from the moment a ten-cycle block of syndrome data is acquired by the software decoding system to the point when the block is fully processed by the decoder during a single experimental run. We observe stable real-time decoding performance with occasional, brief latency spikes from which the system recovers rapidly.}
  \label{fig:latency_vs_block_number_single_run}
\end{figure*}

\begin{figure}
  \centering
  \includegraphics[width=0.85\columnwidth]{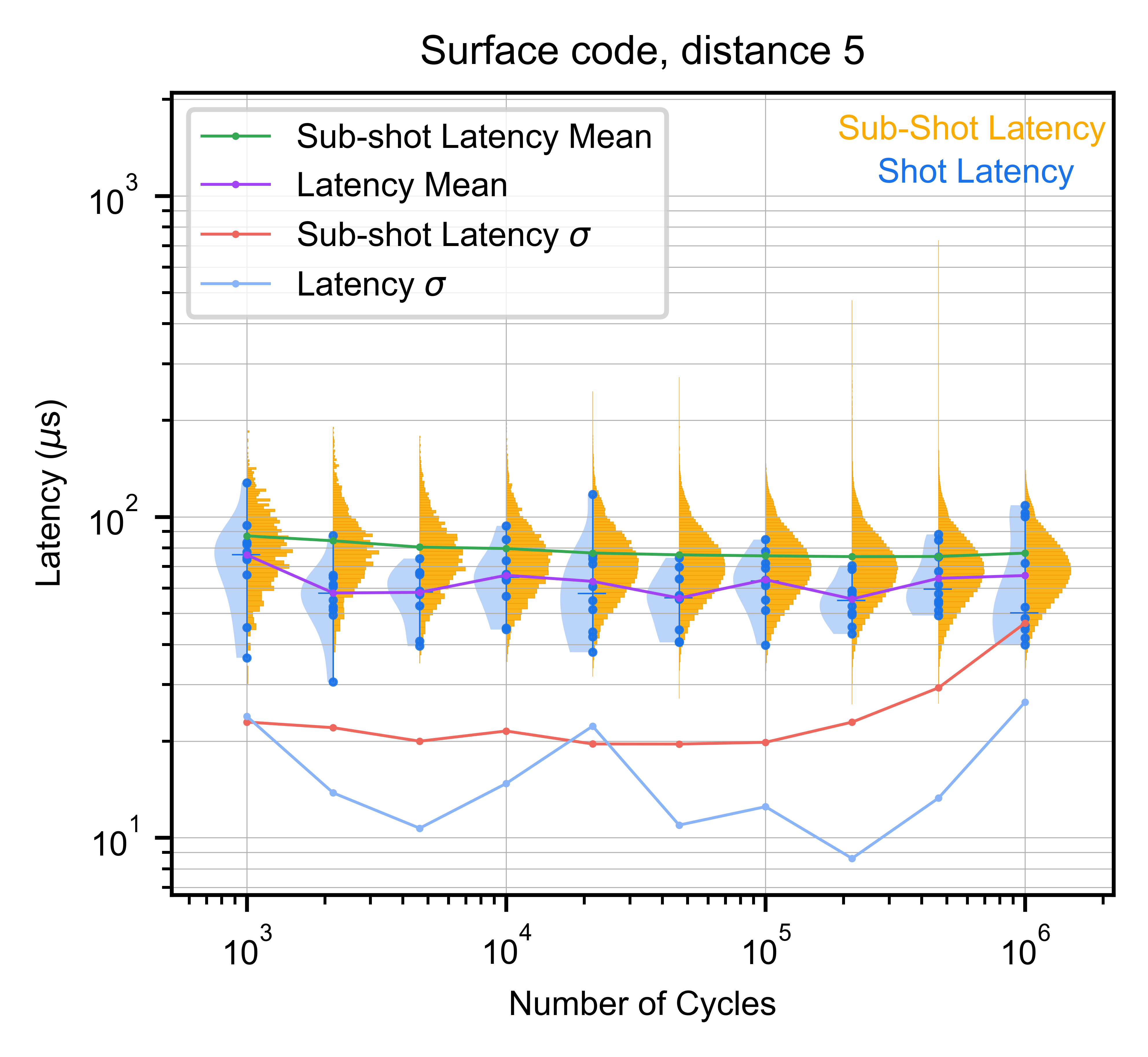}
  \caption{\textbf{Software decoding latency.} Here we show the latency associated with the software components, $T_\text{software}$, as a function of the number of cycles for ten independent experimental shot runs. Blue points indicate per shot latency (10 shots per number of cycles), with horizontal bars denoting the median, and the blue-shaded violin plot illustrating the shot latency distribution. The yellow histograms represent the distribution of sub-shot latencies, obtained by dividing each shot into sub-shots of ten-cycle blocks. Stable latency as a function of the number of cycles indicates the real-time decoding system achieves the required throughput.}
  \label{fig:latency_vs_circuit_depth}
\end{figure}

In order to establish real-time performance of our decoding system we measured $T_\text{software}$ as a function of circuit depth, as shown for a single representative experiment using a quantum circuit with one million cycles in Fig.~\ref{fig:latency_vs_block_number_single_run}. We see that $T_\text{software}$ is roughly constant as a function of time. Moreover, when the latency spikes the system is able to recover quickly. Measurements from many experiments with different number of cycles are shown in Fig.~\ref{fig:latency_vs_circuit_depth}. Stability of $T_\text{software}$ together with evidence that all syndrome information has been analyzed by the decoder, proves that the system processes syndrome information at the rate at which it is generated.

Measurements indicate that in our current real-time decoding system $T_\text{input}< 10$ \textmu s. In a hypothetical system with the addition of $T_\text{output}$ and $T_\text{control}$ in the range of a few microseconds, the reaction time would be within an order of magnitude from that assumed in Ref.~\cite{Gidney2021howtofactorbit}. Improving decoding latency while scaling the system to larger code distances is an outstanding engineering challenge on the path to large-scale fault-tolerant quantum computing.

There are trade-offs between accuracy, latency, and throughput. For example, aggressive batching of syndrome or intermediate information may increase throughput at the expense of increased latency. Similarly, a decoder that short-circuits rare and computationally hard cases of syndrome decoding by making a random guess may achieve higher throughput and lower latency at the expense of reduced accuracy. The requirements interact in other ways, too. For example, in early fault-tolerance experiments, the logical circuit depth budget may be increased either by improving decoding accuracy or by reducing decoding latency.

A different kind of trade-off exists in the fundamental choice between decoders implemented in hardware and decoders implemented in software. The former promise greater performance while the latter offer greater flexibility. Our experiments demonstrate that a parallel and optimized C++ software decoder can achieve real-time performance when provided exclusive access to a subset of CPUs. The flexibility of software-based decoding facilitates ongoing research in decoding algorithms and fault-tolerant constructions. In particular, software syndrome decoding can handle broken or disabled components \cite{Nagayama_2017}, allows easing of hardware requirements \cite{McEwen2023relaxinghardware}, and promises support for novel and customized constructions for early fault-tolerant quantum algorithms.

\subsubsection{Correlated Parallel Blossom}

In this section we describe our real-time decoder, a multi-threaded implementation of a correlation-augmented minimum-weight perfect matching decoder capable of processing a stream of syndrome data, using a matching graph buffer that does not scale with the number of cycles. 
Our decoder uses Sparse Blossom~\cite{higgott2023sparse} as the matching engine (main subroutine) and upgrades it to an accurate real-time decoder by making the following major changes:

\paragraph{Preweight correlations.}\label{sec:preweight_corr}

Sparse Blossom, as described in Ref.~\cite{higgott2023sparse}, assumes that the noise model is graphlike (each error causes at most two detection events) and ignores ``hyperedge'' error mechanisms that can cause more than two detection events, such as $Y$ errors in the CSS surface code.
We improve the accuracy of Sparse Blossom by reweighting the graph using a ``preweight correlations'' subroutine.
This is similar to the reweight strategy introduced in Ref.~\cite{Paler2023pipelinedcorrelated}, which itself is a fast approximation of the two-pass correlation strategy introduced in Ref.~\cite{fowler2013optimalcomplexitycorrectioncorrelated}.
Concretely, we iterate over the detection events and identify any edge satisfying either of the following conditions:
\begin{enumerate}
    \item The edge has a detection event at both of its endpoints.
    \item The edge is a boundary edge that has a detection event $v$ at its endpoint, and furthermore there are no detection events directly adjacent to $v$.
\end{enumerate}
If an edge $e$ satisfies either of these conditions then we add it to the set $C$ of initially chosen edges.
For each edge in $C$ we then condition on an assumption that the associated error mechanism occurred to obtain a posterior probability that other related errors occurred, updating edge weights accordingly.
For example, if an edge $e$ in $C$ corresponds to an $X$ error at a particular location of a CSS surface code circuit, then from our knowledge of the noise model we know that a $Y$ error could have occurred at the same location and hence we lower the edge weight corresponding to a $Z$ error at the same circuit location.

\paragraph{Block-based parallelization.}

We parallelize Sparse Blossom following a similar strategy to how Fusion Blossom parallelizes Parity Blossom~\cite{wu2023fusionblossom}.
Our decoder partitions the matching graph into \textit{blocks}, where each block consists of $M$ syndrome measurement cycles.
Each block is assigned to a thread, which first reweights the graph using the preweight correlations strategy described in Section\,\ref{sec:preweight_corr}.
The same thread then runs Sparse Blossom on the block, with the Sparse Blossom timeline processed until every region is matched either to another region, the boundary of the graph, or the boundary of the block.
Similar to Fusion Blossom~\cite{wu2023fusionblossom}, neighbouring blocks are then \textit{fused} in stages, again using Sparse Blossom as the matching engine.

The fusion graph we use is different from the ones proposed in Fusion Blossom.
Specifically, we apply two layers of fuses (our fusion graph has height three): we first fuse each 
even block with the odd block immediately following it, and then we fuse each odd block with the even block immediately following it.
Here, a block is even or odd if its sequential index is even or odd, respectively.
Rather than adding a root node to the fusion graph, we instead declare a heralded error if any regions remain matched to a block boundary once fusing has completed.
We ensure that the rate of heralded failures is negligible relative to other contributions to the logical error rate by choosing a sufficiently large number of cycles per block $M$.
We find it is sufficient to set $M=10$ for the distance-3 and distance-5 surface codes and $M=90$ for the distance-29 repetition code.

Finally, we extract the predicted logical observable from each block as a bitmask and undo the edge reweights.
The overall prediction of the decoder is then the bitwise XOR of the predicted logical observable bitmask for each block.

\paragraph{Constant-sized graph buffer.}

Rather than storing the full matching graph in memory, we instead use a constant-sized graph buffer.
This graph buffer emulates the full matching graph while only storing 128 contiguous blocks at any instant.
A separate \textit{grapher} thread maintains the graph buffer, ensuring that the buffer always contains the region of the graph being acted on by the decoder.
The grapher exploits the repeating pattern of the memory experiment matching graph and does not rewrite any of the graph buffer in the bulk of the experiment, with the only significant work being done at the beginning and end of the experiment where the local structure of the graph changes.

%% file: text_sm/error_budget_sims.tex
\subsubsection{Surface Code Simulation Details}
\label{sec:sim_details} 
We briefly summarize the methods and high-level workflow used to simulate the error correction experiments. Except when otherwise noted, the simulation details are equivalent to those discussed in Ref.~\cite{google2023suppressing}. The gate sequence matching each memory experiment is first represented in a \texttt{cirq.Circuit}~\cite{cirq_developers_2024_11398048}. The circuit is then dressed with additional channels corresponding to different error mechanisms. Each error channel is represented numerically using Kraus operators, which are compiled specifically for each noise model and qubit based on experimental characterizations. The following mechanisms are included:
\begin{itemize}
    \item Decoherence, quantified by decay ($T_1$), pure dephasing ($T_{\phi}$), and passive heating of qubits to state $\ket{2}$.
    \item Readout and reset error.
    \item Dephasing-induced leakage of the higher frequency qubit during the CZ gate. (This is represented by a transition from state $\ket{11}$ to $\ket{02}$.)
    \item Stray coupling crosstalk between nearest-neighbor or diagonal-neighbor qubits during parallel CZ operation.
    \item Excess error (not accounted for by previous sources) on single-qubit and CZ gates, as well as during idling of data qubits during qubit measurement and reset.
    \item Transport of leakage between CZ-gate qubits through higher-excitation transitions (e.g., $\ket{12}\rightarrow\ket{30}$).
\end{itemize}
Beyond these errors already included in Ref.~\cite{google2023suppressing}, we also allow for imperfect operation of the data-qubit leakage removal operation (DQLR)~\cite{Miao2023}. For simplicity, this is modeled by a phenomenological model matching experimentally observed reset fidelity for state $\ket{2}$. The Kraus operators of this channel are given by

\begin{align}
\label{Suppl:Kraus-DQLR} 
    K_{ij} & = \sqrt{P_{j\rightarrow i}} \ketbradt{i}{j} \\
    K_0 & = \sqrt{I - \sum_{i,j} K_{ij}^\dagger K_{ij}}
\end{align}
where $P_{j\rightarrow i}$ is the probability that standard basis state $\ket{j}$ of the data qubit is reset to state $\ket{i}$.
Notably, this channel acts trivially on the computational subspace. 

\begin{figure}[t!]
    \centering
    \includegraphics[width=\columnwidth, trim =0cm 0cm 0cm 0cm,clip=true]{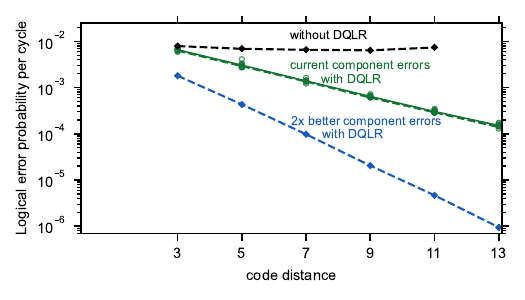}
    \caption{\textbf{Simulated logical error rates for larger code distances.} For a given code distance, the green open circles indicate the LERs of ten different non-uniform error models obtained by re-sampling (bootstrapping) the measured component error probabilities of our distance-5 surface code on the 72-qubit processor. Note that the circles are overlapping. The green solid line denotes the geometric mean of the LERs indicated by the green circles. The dashed lines depict LERs for error models that, except for stray crosstalk errors, are uniform over each qubit. The dashed green and black lines respectively show the LERs with and without data qubit leakage removal (DQLR) gates at the end of each error correction cycle. Finally, the dashed blue line shows the logical error rates expected for a device with twice better component error probabilities and with DQLR.}
    \label{fig:suppl_scalability_plot}
\end{figure}

After the circuit is dressed with all relevant error channels, we apply a Generalized Pauli Twirling Approximation to each noise channel. This converts each noise channel to a generalized Pauli channel that also includes leakage, thereby making it compatible with Clifford simulation methods. Finally, the noisy circuit is parsed and passed to the \texttt{Pauli+} simulator~\cite{google2023suppressing}, which generates samples of simulated experimental data.

\subsubsection{Surface Code Performance at Large Code Distances}
\label{sec:performance_of_larger_surf_codes}
In this section, we simulate the performance of surface codes at code distances larger than those realized experimentally. In particular, we discuss (a) the effect of non-uniform component errors on the logical error rate (LER) and the error suppression factor $\Lambda$; (b) the achievable logical error rates if we improve our components' fidelity by a factor of two; (c) the effect on the exponential suppression of the logical error rate with code distance $d$ if we forego DQLR; and (d) the crossover regime where we can observe finite-size error suppression above threshold. We point out that, for efficiency, all results of this section (except when noted otherwise) are obtained using a correlated minimum-weight perfect matching decoder~\cite{google2023suppressing} that is less accurate than the one used in the main text. Moreover, we carry out the simulations using the component error parameters of those of the 72-qubit processor.

\paragraph{Effect of non-uniform component errors on surface code performance.} 

The component error probabilities of our devices are intrinsically non-uniform. To investigate the effect of the non-uniformity of component errors on the performance of surface codes at large code distances, we generate synthetic non-uniform error models by resampling (bootstrapping) the measured component errors of the $d=5$ surface code grid. For codes of each distance $d = 3, 5, 7, 9, 11$ and 13, we generate ten statistically independent synthetic error models and compute the corresponding LERs.

The results are depicted by the open green circles in Fig.~\ref{fig:suppl_scalability_plot}. The solid green line joins the geometric means of the computed LERs. For comparison, we also perform surface code simulations where, for each error type, we homogenize the component errors by taking the arithmetic mean. In these simulations, however, the stray coupling crosstalk errors are still non-uniform, since they depend on the qubit frequencies. The LERs for these (almost) uniform error models are shown by the dashed green line in Fig.~\ref{fig:suppl_scalability_plot}. We find that these LERs show good agreement with the geometric means of the LERs from the non-uniform error models. Both dashed and solid green lines indicate that logical errors are exponentially suppressed at large code distances with a suppression factor $\Lambda\approx2$. From the solid green line, we have $\Lambda_{3/5} = 2.16, \Lambda_{5/7} = 2.2, \Lambda_{7/9} = 2.16, \Lambda_{9/11} = 2.12$ and $\Lambda_{11/13} = 2.01$, where $\Lambda_{d/d+2} \equiv {\rm LER}(d)/{\rm LER}(d+2)$. We perform all these calculations with DQLR gates that remove leakage states $|2\rangle$ and $|3\rangle$ with 100\% and 50\% fidelities, respectively. In Eq.~\eqref{Suppl:Kraus-DQLR}, this corresponds to nonzero transition probabilities $P_{2\rightarrow 1}=1, P_{3\rightarrow 2}=0.5, P_{3\rightarrow 1}=0.5$. These results show that the error non-uniformity of our surface code components induces a relatively small spread in the logical error rates at large code distances. Additionally, the observed logical error suppression factor  $\Lambda_{3/5} \gtrsim 2$  holds at larger code distances, despite the non-uniformity of component errors. 

\paragraph{Performance forecast for devices with twice better component errors.} 
By fitting the data of the solid green line of Fig.~\ref{fig:suppl_scalability_plot} with ${\rm LER}(d)= C\cdot[1/\Lambda]^{(d+1)/2}$, we obtain $\Lambda \approx 2.14$ and coefficient $C\approx3.0\times10^{-2}$. This implies that to achieve a logical error rate of, e.g., ${\rm LER}=10^{-6}$ would require a device that could hold a surface code with a minimum distance $d=27$ (1457 qubits). An alternative path to achieving a ${\rm LER}$ of $10^{-6}$ is to continue improving the surface code component fidelities. We perform a simulation to forecast the performance of surface codes with component errors that are twice better than what we have demonstrated in this work. The results  are indicated by the blue dashed line in Fig.~\ref{fig:suppl_scalability_plot}. The  predicted logical error suppression factor is $\Lambda\approx 4.5$ and a $10^{-6}$ logical error rate can be achieved with a $d=13$ surface code (337 qubits). As shown in Section~\ref{sec:Lambda-budget}, the most important component errors to improve are CZ errors and data qubit idle errors. 

\paragraph{Importance of DQLR in surface codes.}
In this section, we demonstrate through numerical simulations the importance of the DQLR operation for logical device performance. In Ref.~\cite{Miao2023}, it was shown that data qubit leakage removal is instrumental to achieving the exponential suppression of logical errors with code distance when physical errors are well below the error threshold. The reason for this is that leakage states (e.g. $|2\rangle$, $|3\rangle$, etc.) that survive over several error correction cycles create time correlations in error detection events that confuse the error decoder. DQLR limits the growth of leakage populations and the lifetime of leakage states (ideally to one cycle), recovering the exponential logical error suppression with code distance promised by quantum error correction. 

The green and black dashed lines of Fig.~\ref{fig:suppl_scalability_plot} show the logical error rate as a function of code distance with and without DQLR, respectively. We use component errors that are uniform, except for the errors due to stray coupling crosstalk. The error parameters used in these simulations correspond to experimental values of the 72-qubit processor. Without DQLR, we find that the logical error suppression factor $\Lambda_{d/d+2}$ saturates to a value around 1 at large code distances, while with DQLR, exponential suppression of logical errors holds to large code distances with a $\Lambda_{d/d+2}$ that does not exhibit a significant dependence on code distance, $d$.

\begin{figure}
    \centering
    \includegraphics[width=\columnwidth]{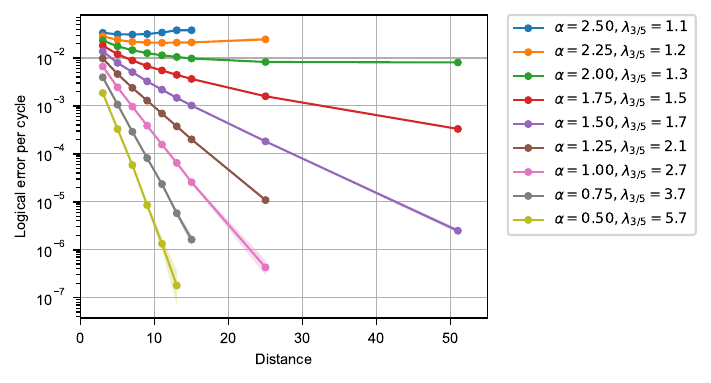}
    \caption{\textbf{The crossover regime.} 
    A Pauli simulation based on the 72-qubit device detailed in Fig.~\ref{fig:edcfs} using $3d$ rounds and correlated matching. 
    The factor $\alpha$ scales the error rates of all the Pauli channels uniformly. 
    We observe that for a $\Lambda_{3/5} \leq 1.3$, error suppression eventually saturates with distance, before turning around.
    However, for $\Lambda_{3/5} \geq 1.5$, we see the characteristic below-threshold error suppression continuing.}
    \label{fig:crossover}
\end{figure}

\paragraph{The crossover regime.}
It is worth noting that the assumption that $\Lambda$ is distance-independent can break down at $\Lambda_{3/5} \approx 1$, as was reported in Ref.~\cite{google2023suppressing}.
This can occur even in a simple Pauli simulation, where one can achieve error suppression at finite sizes above threshold that will eventually turn around.
To illustrate this crossover regime, we employ a simple Pauli simulation of the 72-qubit processor, see Fig.~\ref{fig:crossover}.
There, we observe that for $\Lambda_{3/5} \leq 1.3$, the initial error suppression diminishes rapidly with distance.
We note that the threshold is much better behaved when reporting the error rate per $d$ rounds~\cite{stephens2014fault}, which is relevant to the logical operation of the surface code.

\subsubsection{Surface Code $1/\Lambda$ Error Budget}
\label{sec:Lambda-budget} 
We construct a $1/\Lambda$ error budget for our 72-qubit processor in a manner similar to Ref.~\cite{google2023suppressing}. We write $(\Lambda_{3/5})^{-1}$ as a sum of contributions from each error channel $i$ that is considered in the simulation:

\begin{align}
(\Lambda_{3/5})^{-1} = \sum_i w_i \, p_{\rm expt}^{(i)}
\label{eq:1_Lambda_budget_formula}.
\end{align}

\begin{table*}[htb!]
\centering
    \begin{tabular}{|l|c|c|c|}
    \hline
      \multicolumn{1}{|c|}{Component}   & $p_{\rm expt}^{(i)}$ & $w_i$ & $1/\Lambda$ contrib.\\   \hline
      CZ gates (doesn't include CZ crosstalk \& CZ leakage)\hspace{0.5em} & \hspace{0.5em}$2.8\!\times\!10^{-3}$\hspace{0.5em} & \hspace{0.5em}65\hspace{0.5em} & \hspace{0.5em}0.182 (41\%)\hspace{0.5em}  \\
      CZ crosstalk & $5.5\!\times\!10^{-4}$& 91 & 0.05 (11\%)\\ 
      CZ leakage & $2.0\!\times\!10^{-4}$ & 108 & 0.022 (5\%)\\
      Data qubit idle & $0.9\!\times\!10^{-2}$ &10 & 0.09 (20\%) \\
      Readout & $0.8\!\times\!10^{-2}$ & 6 & 0.048 (11\%) \\
      Reset & $1.5\!\times\!10^{-3}$ & 6 & 0.009 (2\%) \\
      SQ gates & $6.2\!\times\!10^{-4}$ & 63 & 0.039 (9\%)  \\
      Leakage (heating) & $2.5\!\times\!10^{-4}$ & 18 & 0.005 (1\%)\\
      \hline
    \end{tabular}
    \caption{\textbf{Error budget calculation for $(\Lambda_{3/5})^{-1}$} using device parameters from the 72-qubit processor when decoding with correlated matching. We multiply the component errors $p_{\rm expt}^{(i)}$ by the sensitivities (weights) $w_i$ at the error operation point, resulting in the relative contributions to $(\Lambda_{3/5})^{-1}$ shown in the right-most column.} 
    \label{Suppl:1_Lambda_contributions_table}
\end{table*}

The weights $w_i$ are the sensitivities of $(\Lambda_{3/5})^{-1}$ to a uniform increase of the error probabilities of the $i$th error channel from the baseline error operation point. From Table~\ref{Suppl:1_Lambda_contributions_table}, we find that errors related to CZ gates (the first three rows) contribute about 60\% of the $1/\Lambda$ budget. The next important contributor to $1/\Lambda$ is data qubit idle errors (about 20\%), followed by readout and reset errors (12\%) and single-qubit gate errors (9\%). We remark that the sensitivities ($w_i$) are obtained assuming perfect DQLR and using a correlated matching decoder for efficiency~\cite{fowler2013optimalcomplexitycorrectioncorrelated}. The prediction from  Eq.~\eqref{eq:1_Lambda_budget_formula} and Table~\ref{Suppl:1_Lambda_contributions_table} yields $\Lambda_{3/5}=2.25$, while directly calculating the ratio of logical error rates from simulated distance-3 and distance-5 codes yields $\Lambda_{3/5}=2.17$.

%% file: text_sm/break_even.tex
The average channel fidelity of a quantum channel $\cal E$:~$\rho \to {\cal E}(\rho)$ to a target identity channel is defined as
\begin{align}
    \overline{\cal F}[{\cal E}] = \int d\psi \langle \psi | {\cal E}(|\psi\rangle\langle\psi|) |\psi\rangle, \label{average_gate_fidelity}
\end{align}
where the integral is over the uniform measure on the state space, normalized so that $\int d\psi=1$. This fidelity characterizes the closeness of a quantum channel to the identity without giving preference to any particular axis on the Bloch sphere. 

As shown in Ref.~\cite{nielsen2002simple}, the uniform averaging in Eq.~\eqref{average_gate_fidelity} is equivalent to averaging over the six cardinal points on the qubit Bloch sphere, the eigenstates of the Pauli operators,
\begin{align}
    \overline{\cal F}[{\cal E}] = \frac{1}{12}\sum_{P=X,Y,Z}{\rm Tr}\left[P{\cal E}(P)\right] + \frac12,
    \label{nice_formula}
\end{align}
which becomes evident by using the linearity of the quantum channels, ${\cal E}(P)={\cal E}(|{+P}\rangle\langle{+P}|) - {\cal E}(|{-P}\rangle\langle{-P}|)$.
Eq.~\eqref{nice_formula} provides an experimental protocol for extracting fidelity of an {\it arbitrary} channel to the identity.

The simplest way to implement an identity is via an idling operation.
For the physical qubits subject to amplitude damping at rate $\gamma_1$ and white-noise dephasing at rate $\gamma_\varphi=\gamma_2-\gamma_1/2$, idling for time $t$ results in
\begin{align}
    \overline{{\cal F}}(t) = \frac{e^{-\gamma_1 t}+2e^{-\gamma_2 t}}{6} + \frac{1}{2}. \label{fidelity_physical}
\end{align}
On the other hand, for a qubit subject to Pauli channel with Pauli error rates $\gamma_X$, $\gamma_Y$ and $\gamma_Z$, we have:
\begin{align}
    \overline{{\cal F}}(t) = \frac{e^{-2(\gamma_Y+\gamma_Z) t} + e^{-2(\gamma_X+\gamma_Z) t} + e^{-2(\gamma_X+\gamma_Y) t}}{6}  + \frac{1}{2} \label{fidelity_logical} .
\end{align}

Hence, qubits subject to different error channels experience different decay of fidelity in time, which complicates their direct comparison. However, at short times this decay can be approximated as 
\begin{align}
    \overline{\cal F}(\delta t) = 1 - \frac{1}{2}\Gamma \delta t,
\end{align}
where $\Gamma$ is an effective depolarization rate and $1/\Gamma$ is an effective lifetime of a uniformly sampled pure state. This quantity can be directly compared among qubits with different error channels. Note that this metric does not favor qubits with extremely biased noise \cite{grimm2020stabilization,google2021exponential,reglade2024quantum}, for which $\Gamma$ is dominated by the largest error rate. 
For physical qubits, from Eq.~\eqref{fidelity_physical} we obtain
\begin{align}
    \Gamma_{\rm physical} = \dfrac{\gamma_1 + 2\gamma_2}{3}.
\end{align}
For a qubit encoded in the surface code, we model a logical Pauli channel~\cite{bravyi2018correcting, sivak2023real} with error probabilities $p_X$, $p_Y$ and $p_Z$ per cycle. When these error probabilities are small, the channel can be modeled as continuous in time, with rates $\gamma_X\approx p_X/t_c$, $\gamma_Y\approx p_Y/t_c$, and $\gamma_Z\approx p_Z/t_c$, where $t_c$ is the cycle duration. For such a continuous logical Pauli channel, from Eq.~\eqref{fidelity_logical} we obtain
\begin{align}
    \Gamma_{\rm logical} = \frac{4}{3}(\gamma_X + \gamma_Y+\gamma_Z)
\end{align}

Note that to first order $p_X+p_Y+p_Z=2\varepsilon_d-p_Y$, with $\varepsilon_d$ defined in the main text as the logical error probability per cycle averaged over the $X$ and $Z$ eigenstates. This equation allows us to establish a strict upper bound on $\Gamma_{\rm logical}$, since $p_X+p_Y+p_Z\leq2\varepsilon_d$, resulting in 
\begin{align}
    \Gamma_{\rm logical} \leq \dfrac{8\varepsilon_d}{3t_{\rm c}}.
\end{align}
Simulations suggest \cite{gidney2024inplace, gidney2023yoked} that in the surface code $p_Y\sim p_X p_Z$ is suppressed compared to $p_X$ and $p_Z$, because $Y$ errors predominantly occur as independent $X$ and $Z$ errors. Hence, we expect that this bound is also tight.
One illustrative consequence of this is that the lifetime of $Y$ eigenstates in the surface code would be approximately twice shorter than the $X$ and $Z$ eigenstates, although due to a more complicated preparation~\cite{gidney2024inplace}, we do not measure this lifetime in our experiment.

Following Refs.~\cite{ofek2016extending, ni2023beating, sivak2023real}, we define the gain $G$ as an improvement of the effective lifetime of an error-corrected logical qubit over the best physical qubit in the same system (with the break-even point corresponding to $G=1$). This metric represents the usefulness of QEC in protecting quantum information in time; however, it does not take into account the cost of state preparation, measurement, or gates. 

In our system, comparing against the qubit with the longest measured physical lifetime, we have
\begin{align}
    1/\Gamma_{\rm physical} &= 119\pm13\,\text{\textmu s}, \\
    1/\Gamma_{\rm logical} &= 291\pm6\,\text{\textmu s},
\end{align}
which leads to $G=2.4\pm0.3$. This demonstrates a beyond break-even quantum memory realized with a multi-qubit code, extending previous results with bosonic codes \cite{ofek2016extending, ni2023beating, sivak2023real}.

%% file: text_sm/rep_code_floor.tex
\begin{figure}
    \centering
    \includegraphics{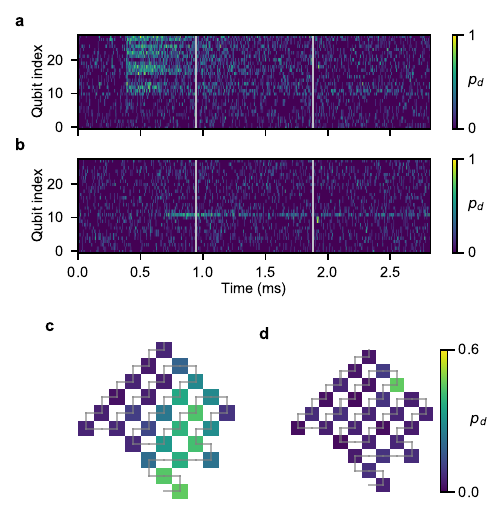}
    \caption{\textbf{Repetition code.} \textbf{a,} Detection probability as a function of time for the measure qubit corresponding to each detector (indexed along repetition code direction), during a burst event. Detections are smoothed with a $\sigma=2$ Gaussian filter. Vertical lines represent boundaries between shots. \textbf{b,} Same as \textbf{a}, but for the case of a single noisy detector. \textbf{c} and \textbf{d,} Detection probability averaged over the first ~200 ns of the burst event and single noisy detector event, respectively, plotted on the qubit grid. Grey lines represent the repetition code gate direction.}
    \label{fig:repetition_supplement}
\end{figure}

In the distance-29 repetition code experiment, we achieve $\Lambda>8$ and observe an apparent logical-error-per-cycle floor of $10^{-10}$ at large distances. This result constitutes a four orders-of-magnitude improvement compared to the lowest logical error per cycle achieved in previous experiments \cite{google2021exponential}. However, even this lower error rate floor will be limiting for larger fault-tolerant circuits. We find that the logical errors at high distances are caused by rare events with distinct detection event probability signatures. In this section, we report relevant metrics for these rare events.

One type of event presents as a correlated burst of detection event probability, with a sharp rise followed by an exponential decay to baseline levels over the course of several shots. An example is given in Fig.~3d, and another is presented in Fig.\,\ref{fig:repetition_supplement}. These events account for all the distance-27 logical errors we observe, and half of the distance-21 to distance-25 errors. The measure qubits corresponding to the firing detectors are grouped spatially, not necessarily in sequence with the snaking repetition code order. The bursts occur roughly once per hour, or once every $3 \times 10^6$ shots, in both bit- and phase-flip codes. The decay timescale from fitting the detection fractions to a decaying exponential is around 400-700 \textmu s. This is in stark contrast to the detection fraction bursts caused by quasiparticles observed in previous work \cite{McEwen2022, google2023suppressing}, which happened once every $10$~seconds with a recovery time of $25-30$~ms. 

The other type of limiting event presents as a single noisy detector, whose likelihood of firing rapidly increases and stays high for 1-2~ms. This event causes a high-distance (defined here as $d>19$) error also about once per hour of data acquisition. The detector which is noisy varies from event to event. An example is presented in Fig.\,\ref{fig:repetition_supplement}.

%% file: text_sm/qecx_details.tex
\begin{figure*}[t!]
    \centering
    \includegraphics[width=1.95\columnwidth]{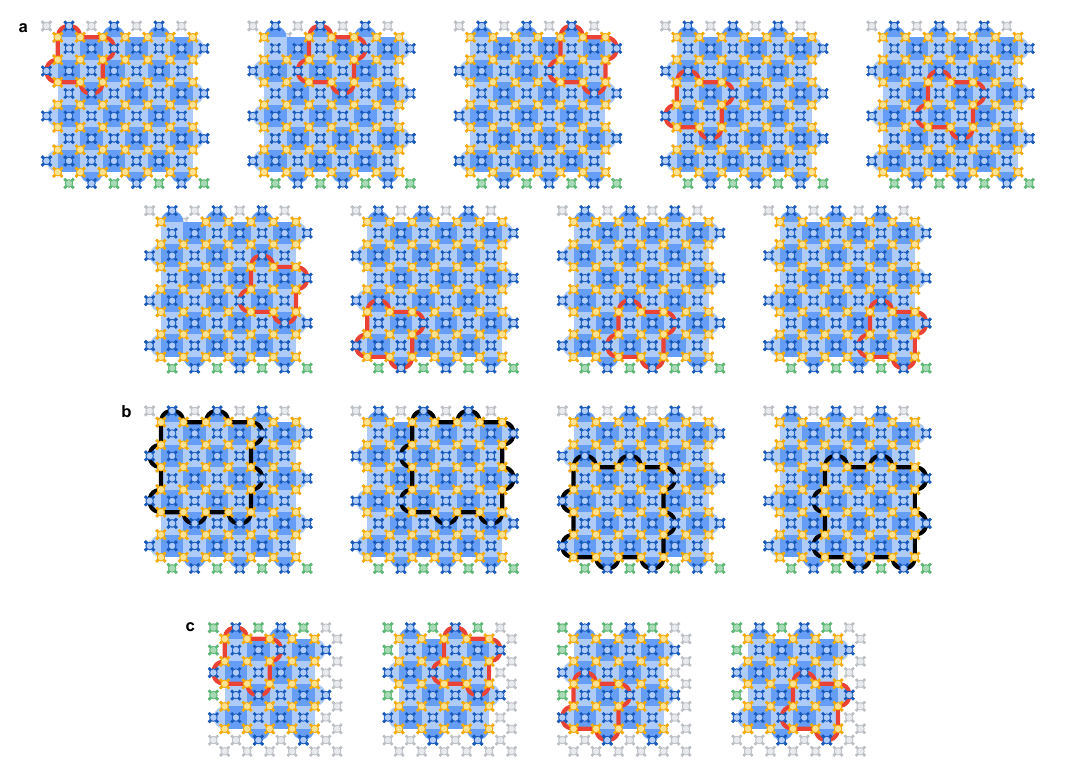}
    \caption{\textbf{Surface code grid layout details.} \textbf{a,} Layout of distance-3 grids used on the 105-qubit processor. \textbf{b,} Layout of distance-5 grids used on the 105-qubit processor. \textbf{c,} Layout of distance-3 grids used on the 72-qubit processor.}
    \label{fig:sub_grids_supplement}
\end{figure*}

\begin{figure}[htb!]
    \includegraphics{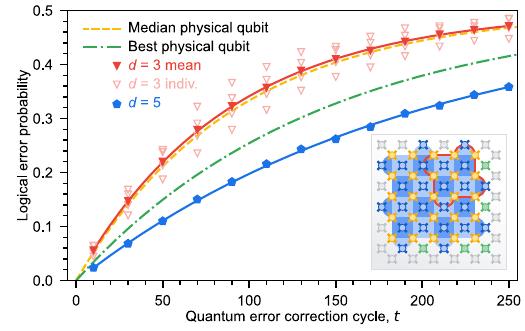}
    \caption{
    \textbf{Logical performance on the 72-qubit processor.}
    Here we show a similar result to Fig.~1c of the main text, except on the 72-qubit processor. Decoding is performed with the neural network.
    The neural network decoder achieves $\varepsilon_5 = 0.252\% \pm 0.002\%$ and $\Lambda = 2.31 \pm 0.02$.
    It also achieves a break-even memory at distance-5, with a median and best physical qubit lifetime of approximately 73\,\textmu s and 113\,\textmu s, respectively, compared to a logical qubit lifetime of approximately 160\,\textmu s.}
    \label{fig:72q_error_vs_cycles}
\end{figure}

In this section, we present experimental details associated with the qubit grids and quantum circuits used for the surface code experiments.  Fig.\,~\ref{fig:sub_grids_supplement} shows the placement of distance-3 and distance-5 grids on both processors. To compare smaller-distance codes to larger ones, we average over several smaller codes chosen to cover the larger code with minimal overlap, extending the methodology of Ref.~\cite{google2023suppressing}.

\begin{figure*}[htb!]
    \centering
    \includegraphics[width=1.95\columnwidth]{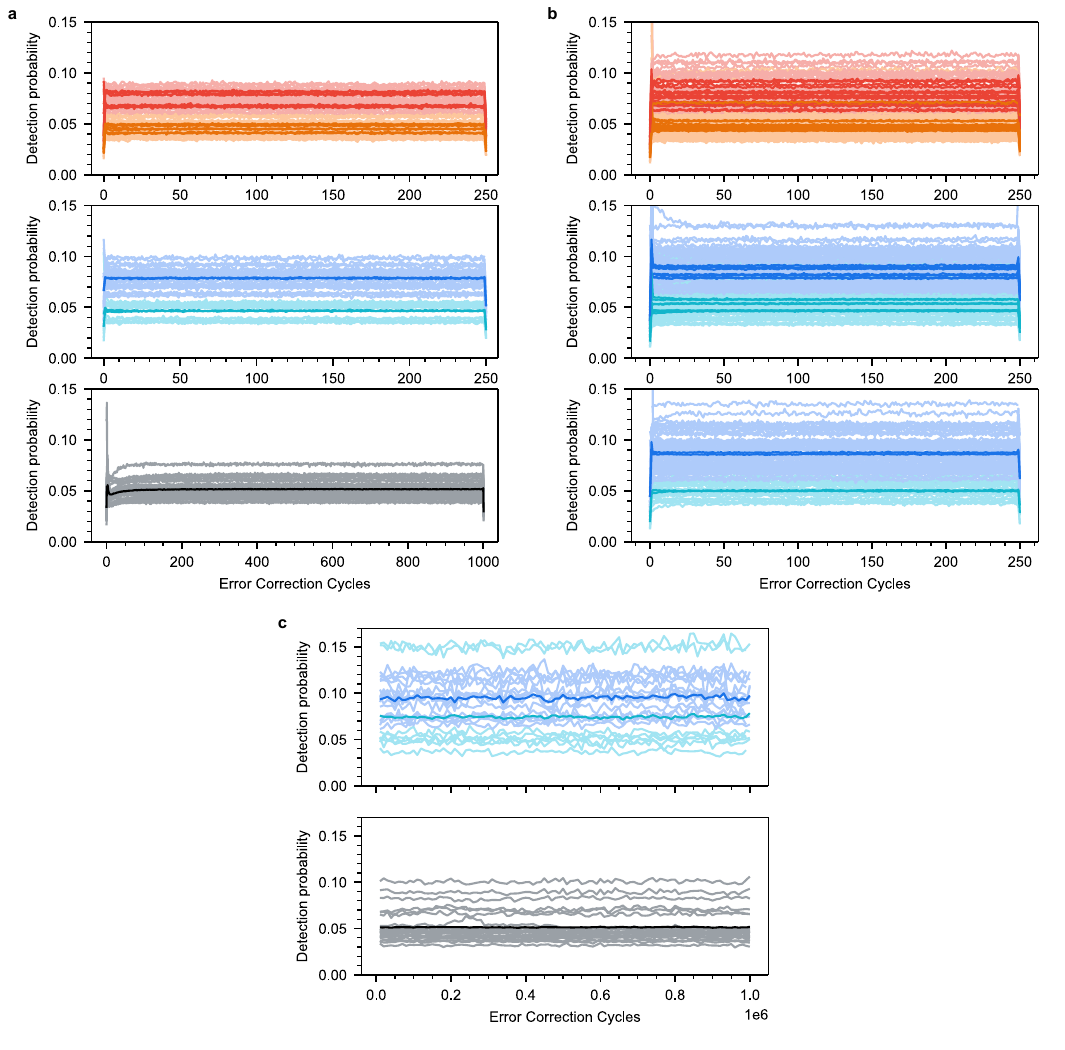}
    \caption{\textbf{Detection event probabilities.} \textbf{a,} Average detection probability as a function of error correction cycles for all detectors in the different grids in the 72-qubit processor. (top) Distance-3 grids, with weight-4 stabilizers in red and weight-2 stabilizers in orange. The average $X$ and $Z$ basis detectors are plotted in bold. (middle) Distance-5 grid, with weight-4 in dark blue and weight-2 in light blue. (bottom) distance-29 repetition code. \textbf{b,} Detection probabilities for different grids in the 105-qubit processor; (top) distance-3 grids, (middle) distance-5 grids, (bottom) distance-7 grid.  \textbf{c,} Detection probability as a function of error correction cycles, run continuously for one million cycles, for (top) the distance-5 grid and (bottom) the distance-29 repetition code in the 72-qubit processor. Here we plot one shot, and apply a rolling average of 10,000 cycles.}
    \label{fig:det_fracs_supplement}
\end{figure*}

\begin{figure*}[htb!]
    \centering
    \includegraphics[width=1.75\columnwidth]{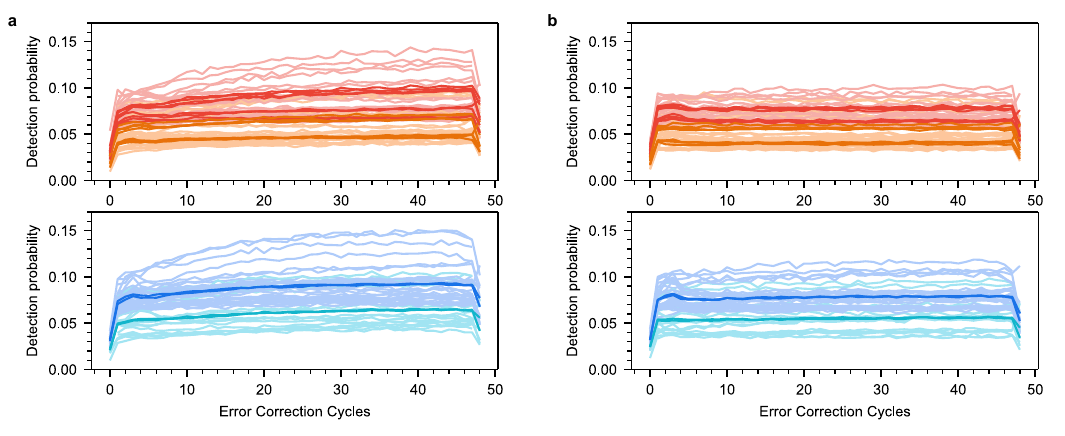}
    \caption{\textbf{Detection event probabilities without and with data qubit leakage removal.} \textbf{a,} Average detection probability as a function of error correction cycles without data qubit leakage removal (DQLR) for all detectors in the distance-3 grids (top) and distance-5 grid (bottom) in the 72-qubit processor. We attribute the rise in detections to leakage accumulation. \textbf{b,} Average detections as a function of error correction cycles with DQLR; the data qubit leakage removal step successfully removes the rise in detections.}
    \label{fig:dqlr_A/B_test}
\end{figure*}

Fig.\,~\ref{fig:72q_error_vs_cycles} shows a representative dataset of logical error versus cycles for the 72-qubit processor taken prior to the time series shown in Fig.\,~2e.
This dataset gives $\Lambda = 2.31 \pm 0.02$.
Fig.\,~\ref{fig:det_fracs_supplement} shows the detection event probabilities from various experiments, including distance-3 and distance-5 codes corresponding to Fig.\,~\ref{fig:72q_error_vs_cycles}, distance-29 repetition codes on the 72-qubit processor, and distance-3, distance-5, and distance-7 codes on the 105-qubit processor. Figure\,~\ref{fig:pauli_simulated_bulk_detection_probabilities} shows the detection probabilities predicted from a Pauli simulation of the 105-qubit device.
We see good agreement, with the simulation predicting detection probabilities for $d=(3,5,7)$ of $p_\text{det}=(8.0\%, 8.5\%, 8.6\%)$ compared to experiment which has $p_\text{det}=(7.7\%, 8.5\%, 8.7\%)$.
The natural rise in detection probability with system size can be ascribed to reducing finite size effects -- for example, each corner, edge, and bulk data qubit is involved in 2, 3, and 4 CZ gates, respectively.
In Fig.\,~\ref{fig:dqlr_A/B_test} we show a comparison of detection probabilities from the data qubit leakage removal (DQLR) test from Fig.~2, noting the rise in detections for the experiments without DQLR, which we attribute to leakage accumulation.

\begin{figure*}[htb!]
    \centering
    \includegraphics[width=1.95\columnwidth]{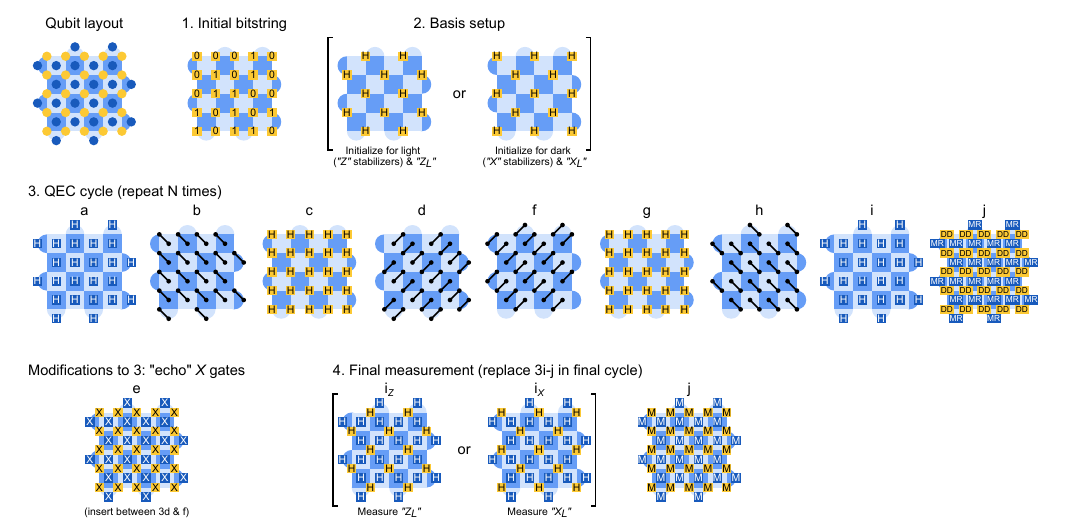}
    \caption{\textbf{Distance-5 surface code circuit.} Spatial layout of data qubits (gold dots) and measure qubits (blue dots) are shown here on a distance-5 grid. 1. We reset all qubits to $\ket{0}$ and then prepare an initial bitstring state on the data qubits using $X$ gates. 2. We apply $H$ (Hadamard) gates to some of the data qubits to convert the initial bitstring (eigenstate of all $Z$ operators) into an eigenstate of half the \textit{ZXXZ} stabilizers, the half matching to logical operator of interest ($X_L$ or $Z_L$) for the specific experiment. Note steps (1) and (2) could be combined into a single moment of Clifford gates, but for simplicity we execute them in two moments as shown. 3. QEC cycle showing the explicit patterns for Hadamard gates and CZ gates as well as dynamical decoupling (DD) and measurement and reset operations (MR). Note that although all stabilizers measure \textit{ZXXZ}, the ``$X$” stabilizers and ``$Z$” stabilizers apply their CZs in a different order (specifically in 3d and f). This pattern is carefully designed to manage ``hook” errors. We modify this gate sequence with additional ``echo” or ``dynamical decoupling” gates, adding $X$ gates to all qubits in 3e (inserted between the middle two CZ moments). 4. For the final measurement, we replace 3i-j with a different Hadamard pattern (transforming the data qubit state to the appropriate basis for logical measurement) and measure all qubits simultaneously (M). This final measurement is used for detectors for the penultimate cycle (measure qubit results) and the final cycle (data qubit results, converted to parities in the relevant basis).}
    \label{fig:qec_circuit_supplement}
\end{figure*}

Fig.\,~\ref{fig:qec_circuit_supplement} outlines the steps of a surface code error correction circuit, broken down into layers of single- and two-qubit gates as well as dynamical decoupling and measurement operations. 

For experiments sweeping over several codes and numbers of cycles, such as Fig.~1c, we interleave experiments of different codes, as in Ref.~\cite{google2023suppressing}. We acquire one (number of cycles, basis, code) dataset at a time, each with 50,000 repetitions (10 initialization bitstrings times 5,000 repetitions). The initialization bitstrings are 5 random bitstrings and their complements; complementary bitstrings have opposite logical eigenvalues for odd-distance surface codes. We shuffle the order in which we measure each number of cycles. The order of enumeration is the following.

\verb|for n in num_cycles:|

\verb|    for basis in (Z, X):|

\verb|        for code in codes:|

\verb|            take_data(n, basis, code)|

%% file: text_sm/uncertainty.tex
\subsection{Logical Performance}
From a statistical point of view, the error suppression factor $\Lambda$ is a random variable, defined with respect to a given collection of codes, whose randomness comes from sampling noise and performance drift for each of the given codes. Evaluating the uncertainty in our estimate of $\Lambda$ involves evaluating the uncertainty in our estimates of the mean logical error per cycle for the collection of codes with a particular distance, and then propagating that through to uncertainty in that scaling with respect to depth, which we detail below.

In Fig.~1c of the main text, we consider logical error probability $p_L$ for a variety of codes for different numbers of error-correction cycles $t$. Each point has statistical uncertainty $\sqrt{p_L(1-p_L)/N}$ with $N=10^5$ repetitions. For example, $p_L=0.1$ corresponds to statistical uncertainty $10^{-3}$.

\begin{table}
\centering
\begin{tabular}{|c|c|c|c|c|}
\hline
\multirow{2}{*}{} & \multicolumn{2}{c|}{$X$ basis} & \multicolumn{2}{c|}{$Z$ basis} \\
\hline
Grid & Error rate & Std. dev. & Error rate & Std. dev. \\
\hline
distance-3 (0, 0) & 0.00561 & 0.00013 & 0.00516 & 0.00008 \\
distance-3 (0, 4) & 0.00562 & 0.00015 & 0.00526 & 0.00014 \\
distance-3 (0, 8) & 0.00519 & 0.00012 & 0.00475 & 0.00021 \\
distance-3 (4, 0) & 0.00672 & 0.00021 & 0.00596 & 0.00023 \\
distance-3 (4, 4) & 0.01044 & 0.00045 & 0.00800 & 0.00030 \\
distance-3 (4, 8) & 0.00641 & 0.00028 & 0.00516 & 0.00039 \\
distance-3 (8, 0) & 0.00678 & 0.00030 & 0.00724 & 0.00024 \\
distance-3 (8, 4) & 0.00982 & 0.00021 & 0.00854 & 0.00012 \\
distance-3 (8, 8) & 0.00507 & 0.00015 & 0.00533 & 0.00014 \\
\hline
distance-5 (0, 0) & 0.00294 & 0.00008 & 0.00229 & 0.00008 \\
distance-5 (0, 4) & 0.00311 & 0.00007 & 0.00238 & 0.00009 \\
distance-5 (4, 0) & 0.00362 & 0.00008 & 0.00326 & 0.00014 \\
distance-5 (4, 4) & 0.00352 & 0.00008 & 0.00309 & 0.00009 \\
\hline
distance-7 (0, 0) & 0.00155 & 0.00004 & 0.00130 & 0.00004 \\
\hline
\end{tabular}
\caption{\textbf{Logical errors per cycle for Fig. 1c.} Both logical error per cycle and standard deviation is shown for each grid and basis using the neural network decoder. Grid labels correspond to the open source~\cite{qec_datasets_zenodo_2024} labeling, e.g. distance-$3$ $(0, 4) \leftrightarrow$ d3\_0+4j. }
\label{tab:fig_errors}
\end{table}

To determine the logical error per cycle $\varepsilon_d$, we fit exponentials to $p_L$ versus number of cycles $t$ (technically, by fitting a line to $\ln\left(1-2p_L\right)$ versus $t$) for each code. We typically see residuals greater than those expected by statistical uncertainty in $p_L$, which we attribute to system performance drift over the course of the experimental sweep. Along with uncertainty in the fitted log slope obtained from linear regression, we use the excess residuals to estimate the uncertainty in $\varepsilon_d$ including this drift. See Ref.~\cite{google2023suppressing} for details. We compute $\varepsilon_d$ and an uncertainty for each code and logical basis and then average over basis and code for data reported in the manuscript. Since the suppression factor $\Lambda$ is defined with respect to the mean logical error per cycle for each distance, we compute the uncertainty in the mean from each individual uncertainty $\sigma_i$, specifically $\sqrt{\Sigma_i \sigma_i^2}/n$. Note this amounts to considering the particular ensemble of codes as fixed, with all the randomness coming from noise in sampling from a given code rather than sampling different codes.

Note that $p_L$, and thus $\varepsilon_d$, depend on the decoding scheme used. In Fig.~1d, we plot $\varepsilon_d$ values for the neural network decoder. The specific values for the neural network decoder and the ensembled synthetic matcher Libra~\cite{jones2024improved} are reported in the left-hand column of Table~\ref{Suppl:table-logical-error-rates}. Furthermore, the specific $\varepsilon_d$ for each grid and basis of the neural network decoder are reported in Table~\ref{tab:fig_errors}.

Similarly, the values corresponding to the accuracy tests on our real-time data in Fig.~4c are reported in the right-hand column of Table~\ref{Suppl:table-logical-error-rates}.
Those experiments included sweeps from 10 to 250 cycles (as in Fig.~1c) but with fewer repetitions ($10^4$ for each code and each number of cycles, split over logical $X$ and $Z$ bases).
The smaller number of repetitions increases statistical uncertainty.
We fit $\varepsilon_d$ for cycles $t > 90$, reserving the shorter experiments for decoder training. The underlying logical error probability data is plotted in Fig.~\ref{fig:real-time-accuracy}.

\begin{figure}
    \centering
    \includegraphics{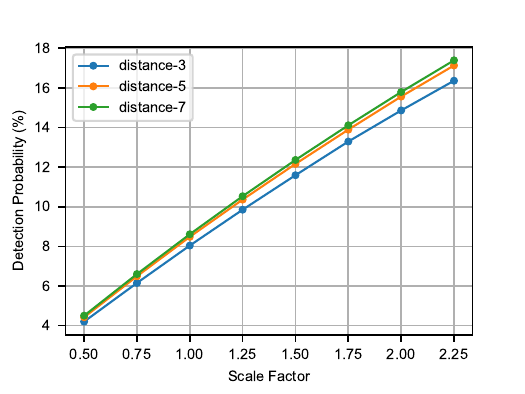}
    \caption{\textbf{Pauli simulated detection event probabilities.} Average detection probability for weight-4 stabilizers predicted from a Pauli simulation of the 105-qubit processor based on Fig.~\ref{fig:edcfs}. Each point is $10^4$ shots of a 100 cycle experiment. Pauli error rates are obtained from the 105-qubit processor. All Pauli errors are scaled uniformly -- 1.0 corresponds to the device Pauli rates. There, for $d=(3, 5, 7)$ we obtain predicted detection probabilities $p_\text{det}=(8.0\%, 8.5\%, 8.6\%)$. We attribute the upwards trend to diminishing finite-size effects. We see reasonable agreement with the experimental detection probabilities $p_\text{det}=(7.7\%, 8.5\%, 8.7\%)$.}   \label{fig:pauli_simulated_bulk_detection_probabilities}
\end{figure}

\begin{figure}
    \includegraphics{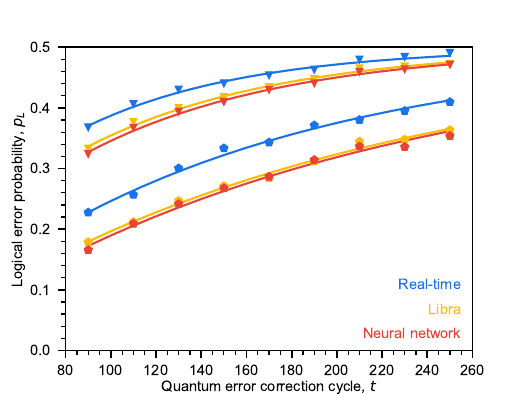}
    \caption{\textbf{Decoder accuracy on the real-time dataset.}
        Underlying logical error probability data for Fig.\,~4c.
        Triangles: distance-3 (averaged over four quadrants and $X$ and $Z$ basis).
        Pentagons: distance-5 (averaged over $X$ and $Z$ basis).
        We compare the accuracy of three different decoding strategies.
        Fits begin from 90 cycles (rather than the usual 10) because these experiments only featured 5,000 shots per individual experiment, rather than the usual 50,000. Consequently, more of the early-cycle samples were used to train the decoder priors.
    }
    \label{fig:real-time-accuracy}
\end{figure}

We compute $\Lambda$ using linear regression of $y = \ln\varepsilon_d$ versus $x = (d+1)/2$, including the uncertainty of the $\varepsilon_d$ values. The slope $m=-\ln\Lambda$, from which we compute $\Lambda=e^{-m}$. Regression gives us uncertainty $\delta m$, and the uncertainty $\delta \Lambda = e^{-m}\delta m$.

We use the same regression analysis for the repetition code $\Lambda$.
In this case, we consider each data qubit idle basis ($X$ and $Z$) separately, as those two cases are sensitive to different physical error mechanisms.
Fitting for $d$ from 5 to 11 (see main text), we obtain $\Lambda_x = 8.27 \pm 0.02$ and $\Lambda_z = 8.55 \pm 0.02$. To report an overall value, we use the mean $\Lambda = (\Lambda_x + \Lambda_z)/2$ and estimate the uncertainty $|\Lambda_x - \Lambda_z|/2$.

For the distance-5 experiments on the 72-qubit processor, we instead consider $\Lambda_{3/5} = \varepsilon_3 / \varepsilon_5$, which is equivalent to the $\Lambda$ fit discussed above for two points.

We compute uncertainties in physical qubit lifetimes by propagating uncertainties in $T_1$ and $T_{2\textrm{CPMG}}$, obtained with \texttt{scipy.curve\_fit}.

\subsection{Logical Error per Cycle From One Point}

\begin{figure}
    \includegraphics{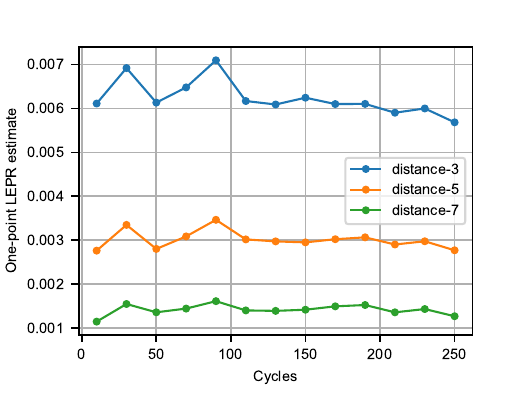}
    \caption{\textbf{Error rates from one-point fits.}
        Here we plot the data from the Fig.~1 of the main text decoded with the neural network.
        For each cycle count $r$, we compute the one-point estimate of the logical error per cycle at that cycle count as $\epsilon_d = 0.5\left(1 - (1 - 2p_L)^{1/r}\right)$, where $p_L$ is the logical error fraction at $r$ rounds, and average over basis and grid.
        We observe that the estimates remain relatively constant for different numbers of cycles, indicating that performance is not degrading for longer cycle counts.}
    \label{fig:one_point}
\end{figure}

In Fig.~2b and 3 of the main text, we estimate the logical error per cycle from experiments with a single number of cycles. The error-injection experiments in Fig.~2b and 3b are 10 cycles long; we choose shorter experiments so we can resolve high logical error $\varepsilon_d \sim 0.1$. The repetition code experiments in Fig.~2a are 1000 cycles so we can resolve $\varepsilon_3 \sim 10^{-3}$ while maximizing experiment length. To estimate the error per cycle $\varepsilon_d$ from a single point $(t, p_L)$, we consider a binomial problem with logical error probability $\varepsilon_d$ at each step. The cumulative error probability $p_L$ after $t$ cycles is the probability of an odd number of logical errors, giving this expression for $p_L$:
\begin{align*}
    p_L &= \frac{1}{2}\left(1 - (1-2\varepsilon_d)^t\right).
\end{align*}
Solving for $\varepsilon_d$,
\begin{align*}
    \varepsilon_d &= \frac{1}{2}\left(1 - (1-2p_L)^{1/t}\right).
\end{align*}
We then propagate statistical uncertainties accordingly. The uncertainties become more noticeable for the low-error experiments (where relatively few errors are observed), such as the higher-distance points in Fig.~3a, where we have a statistical floor based on the total number of cycles in the dataset, $2\times 10^{10}$. We also plot the logical error per cycle for Fig.~1c estimated from one cycle count for different cycles, and then averaged over grid and basis, in Fig.~\ref{fig:one_point}. 